\documentclass[aps,pra,reprint,groupedaddress,showpacs,amssymb]{revtex4-1}
\usepackage{graphicx}
\begin{document}
\title{Bell inequalities under non-ideal conditions}
\thanks{Different format for \cite{Especial12}, same text.}
\author{Jo\~ao N. C. Especial}
\email{joao.especial@imagitec.pt}
\noaffiliation
\date{May 16, 2012}
\begin{abstract}
Bell inequalities applicable to non-ideal EPRB experiments are critical to the interpretation of experimental Bell tests.
In this article it is shown that previous treatments of this subject are incorrect due to an implicit assumption and new inequalities are derived under general conditions.
Published experimental evidence is reinterpreted under these results and found to be entirely compatible with local-realism, both, when experiments involve inefficient detection, if fair-sampling detection is assumed, as well as when experiments have nearly ideal detection and measurement crosstalk is taken into account.
\end{abstract}
\pacs{03.65.Ud, 03.67.Mn}
\maketitle
\section{Ideal EPRB experiments\label{Ideal EPRB experiments}}
The origin of the branch of quantum physics presently concerned with quantum entanglement and non-locality can, historically, be traced to the 1935 article by Einstein, Podolsky and Rosen \cite{EPR35} and from then to the equivalent thought-experiment proposed, in 1951, by Bohm \cite{Bohm51}.

\begin{figure}[ht]
\includegraphics[bb = 148 617 398 717, scale = 0.95, clip]{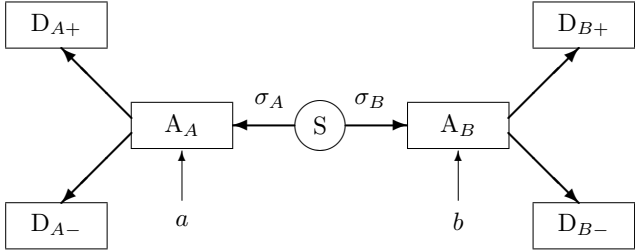}
\caption{Flow diagram of the EPRB experiment with two-channel analyzers.\label{2CEPRBSetup}}
\end{figure}

This thought-experiment, now known as the two-channel analyzer EPRB experiment, involves the setup depicted, as a flow diagram, in Fig.~\ref{2CEPRBSetup}.

The experimental procedure is as follows:
\begin{enumerate}
\item
Pairs of quantum systems, $\sigma_A$ and $\sigma_B$, possibly different, are successively prepared at the source, S, in a non-factorizable quantum state (entangled state);
\item
The two systems are physically separated;
\item
Each quantum system undergoes an independent dichotomic measurement, represented by analyzers $\text{A}_A$ and $\text{A}_B$, parameterized, respectively, by $a$ or $b$.
The measurement outcome is signaled through one of the analyzer's two channels, $+$ or $-$.
The two analyzers need not be symmetric, both in the sense that measurement outcomes in each arm of the experiment may have marginal probabilities $\scriptstyle \neq \frac{1}{2}$ and that the marginal probabilities on both arms of the experiment need not be identical;
\item
Detectors, $\text{D}_{A+}$, $\text{D}_{A-}$, $\text{D}_{B+}$ and $\text{D}_{B-}$, monitor the analyzers' outputs and produce signal detection events, which are recorded.
\end{enumerate}

Ideally, for every pair of prepared quantum systems, two measurement signals would be detected, and thus, two detection events would be recorded, either $+$ or $-$ on the $A$ arm of the experiment, and also either $+$ or $-$ on the $B$ arm.

Under such conditions, the outcome of an experiment run involving $N_{\text{P}}$ pairs of quantum systems, can be summarized by a contingency table as Table~\ref{ContTable}, where the observed frequencies, $f^{AB}(a,b)$, depend on the parameters, $a$ and $b$, and the symbol $\pm$ is used to represent summation over the two possible outcomes, $+$ and $-$.

\begin{table}[ht]
\caption{Outcome summary of an ideal two-channel analyzer EPRB experiment run.\label{ContTable}}
\renewcommand*{\arraystretch}{1.5}
\begin{tabular}{|c|cc|c|}
\hline
$f^{AB}(a,b)$ & $+$ & $-$ & $B$ \\
\hline
$+$ & $f^{AB}_{++}(a,b)$ & $f^{AB}_{+-}(a,b)$ & $f^{AB}_{+\pm}(a,b)$ \\
$-$ & $f^{AB}_{-+}(a,b)$ & $f^{AB}_{--}(a,b)$ & $f^{AB}_{-\pm}(a,b)$ \\
\hline
$A$ & $f^{AB}_{\pm+}(a,b)$ & $f^{AB}_{\pm-}(a,b)$ & $N_{\text{P}}$ \\
\hline
\end{tabular}
\end{table}

In the limit $N_{\text{P}} \to \infty$, the relative frequencies approach probabilities:
\[\forall_{i,j \in \{+,-\}} \colon \qquad p^{AB}_{ij}(a,b) = \lim_{N_{\text{P}} \to \infty} \frac{f^{AB}_{ij}(a,b)}{N_{\text{P}}},\]
and, on their basis, a correlation, $E^{AB}(a,b)$ \cite{Bell64}, can be introduced to describe the degree of association between outcomes on both arms of the experiment:
\[E^{AB}(a,b) := p^{AB}_{++}(a,b) + p^{AB}_{--}(a,b) - p^{AB}_{+-}(a,b) - p^{AB}_{-+}(a,b).\]

This version of the EPRB experiment involves two distinct values of the measurement parameters on either arm of the experiment \cite{CHSH69}, say $a$ and $a'$ on the $A$ arm and $b$ and $b'$ on the $B$ arm.
Let's call $A$ and $A'$ the respective detection records on the $A$ arm and $B$ and $B'$ the similar records on the $B$ arm.

A specific realization of this version of the EPRB experiment can, thus, be fully characterized by the four joint probability tables in Table~\ref{JProbTables}.

\clearpage

\begin{table}[ht]
\caption{Full characterization of an ideal two-channel analyzer EPRB experiment.\label{JProbTables}}
\renewcommand*{\arraystretch}{1.5}
\begin{tabular}{|c|cc|c|}
\hline
$p^{AB}$ & $+$ & $-$ & $B$ \\
\hline
$+$ & $p^{AB}_{++}$ & $p^{AB}_{+-}$ & $p^{AB}_{+\pm}$ \\
$-$ & $p^{AB}_{-+}$ & $p^{AB}_{--}$ & $p^{AB}_{-\pm}$ \\
\hline
$A$ & $p^{AB}_{\pm+}$ & $p^{AB}_{\pm-}$ & $1$ \\
\hline
\end{tabular}
\begin{tabular}{|c|cc|c|}
\hline
$p^{AB'}$ & $+$ & $-$ & $B'$ \\
\hline
$+$ & $p^{AB'}_{++}$ & $p^{AB'}_{+-}$ & $p^{AB'}_{+\pm}$ \\
$-$ & $p^{AB'}_{-+}$ & $p^{AB'}_{--}$ & $p^{AB'}_{-\pm}$ \\
\hline
$A$ & $p^{AB'}_{\pm+}$ & $p^{AB'}_{\pm-}$ & $1$ \\
\hline
\end{tabular}
\begin{tabular}{|c|cc|c|}
\hline
$p^{A'B}$ & $+$ & $-$ & $B$ \\
\hline
$+$ & $p^{A'B}_{++}$ & $p^{A'B}_{+-}$ & $p^{A'B}_{+\pm}$ \\
$-$ & $p^{A'B}_{-+}$ & $p^{A'B}_{--}$ & $p^{A'B}_{-\pm}$ \\
\hline
$A'$ & $p^{A'B}_{\pm+}$ & $p^{A'B}_{\pm-}$ & $1$ \\
\hline
\end{tabular}
\begin{tabular}{|c|cc|c|}
\hline
$p^{A'B'}$ & $+$ & $-$ & $B'$ \\
\hline
$+$ & $p^{A'B'}_{++}$ & $p^{A'B'}_{+-}$ & $p^{A'B'}_{+\pm}$ \\
$-$ & $p^{A'B'}_{-+}$ & $p^{A'B'}_{--}$ & $p^{A'B'}_{-\pm}$ \\
\hline
$A'$ & $p^{A'B'}_{\pm+}$ & $p^{A'B'}_{\pm-}$ & $1$ \\
\hline
\end{tabular}
\end{table}

If outcomes in one arm of the experiment are independent of the choice of measurement parameter made on the other arm, the marginal probabilities will satisfy:
\begin{eqnarray}
p^{AB}_{+\pm} = p^{AB'}_{+\pm} =: p^{A}_{+}, & \qquad & p^{AB}_{-\pm} = p^{AB'}_{-\pm} =: p^{A}_{-}, \label{ApLoc1} \\
p^{A'B}_{+\pm} = p^{A'B'}_{+\pm} =: p^{A'}_{+}, & \qquad & p^{A'B}_{-\pm} = p^{A'B'}_{-\pm} =: p^{A'}_{-}, \label{ApLoc2} \\
p^{AB}_{\pm+} = p^{A'B}_{\pm+} =: p^{B}_{+}, & \qquad & p^{A'B}_{\pm-} = p^{A'B}_{\pm-} =: p^{B}_{-}, \label{ApLoc3} \\
p^{AB'}_{\pm+} = p^{A'B'}_{\pm+} =: p^{B'}_{+}, & \qquad & p^{AB'}_{\pm-} = p^{A'B'}_{\pm-} =: p^{B'}_{-}. \label{ApLoc4}
\end{eqnarray}

These conditions are necessary but, as will be shown in the next section, not sufficient to ensure locality.
Let's say we have 'apparent locality' when these conditions are met.

From the four joint probability tables in Table~\ref{JProbTables}, four correlation values can be obtained:
\[\forall_{X \in \{A,A'\}, Y \in \{B,B'\}} \colon E^{XY} = p^{XY}_{++} + p^{XY}_{--} - p^{XY}_{+-} - p^{XY}_{-+}.\]

The summary statistic for this version of the EPRB experiment, known as the CHSH \cite{CHSH69} Bell-quantity, $S$, is based on these four correlation values and defined as:
\begin{equation}
S := E^{AB} - E^{AB'} + E^{A'B} + E^{A'B'}. \label{SDef}
\end{equation}

Another version of the EPRB experiment results from removing one detector from each arm of the setup in Fig.~\ref{2CEPRBSetup}.
Since only one channel of each analyzer is now monitored, this version is known as the single-channel analyzer EPRB experiment.
Either detector may be removed. Let's choose to keep the $+$ detectors on both arms and remove both $-$ ones.

The summary statistic for this version is known as the CH \cite{CH74} Bell-quantity, $\Delta'$.
This is based directly on the joint and marginal probabilities in Table~\ref{JProbTables} and defined as:
\begin{equation}
\Delta' := p^{AB}_{++} - p^{AB'}_{++} + p^{A'B}_{++} + p^{A'B'}_{++} - p^{A'}_+ - p^B_+. \label{Delta'Def}
\end{equation}

A single-channel analyzer EPRB experiment may also be supplemented with three additional experiments in which one or both of the analyzers are removed and all quantum systems, in the respective arm, directly detected without measurement.
The collection of these four experiments will be designated as the single-channel removable analyzers EPRB experiment.

In this version, the measurement parameters, $a$ and $b$, may be regarded as taking three alternative settings on either arm, say $a$, $a'$ and $\infty$ on the $A$ arm and $b$, $b'$ and $\infty$ on the $B$ arm, where the $\infty$ setting stands for removal of the analyzer.
Let's designate by $A$, $A'$ and $A''$ the respective detection records on the $A$ arm and by $B$, $B'$ and $B''$ those on the $B$ arm.

The summary statistic for this version is the CH \cite{CH74} Bell-quantity $\Delta$ defined as follows:
\begin{equation}
\Delta := p^{AB}_{++} - p^{AB'}_{++} + p^{A'B}_{++} + p^{A'B'}_{++} - p^{A'B''}_{++} - p^{A''B}_{++}, \label{DeltaDef}
\end{equation}

The maximum and minimum values of this quantity can be combined into another summary statistic known as the Freedman \cite{FC72} Bell-quantity, $\delta$.
This is defined as:
\begin{equation}
\delta := \frac{\Delta_{\text{max}} - \Delta_{\text{min}}}{4}. \label{deltaDef}
\end{equation}

The probabilities in Table~\ref{JProbTables} and those in its generalization for the removable analyzers version, are constrained by their non-negativity, by their total of 1 in each sub-table and, under apparent locality, by equalities~(\ref{ApLoc1})-(\ref{ApLoc4}).
All these constraints are linear on the probabilities.

The four Bell-quantities, $S$, $\Delta'$, $\Delta$ and $\delta$, are also linear on the probabilities and, thus, the bounds that those constraints place on these quantities can be established by solving a minimization and a maximization linear programming problem for each of the quantities, taken as objective function.

Solving these problems, using, for instance, the simplex method, establishes the following bounds, applicable to ideal EPRB experiments under apparent locality, as defined above:
\begin{eqnarray}
-4 \leqslant & S & \leqslant 4, \label{ApLocB1} \\
-\frac{3}{2} \leqslant & \Delta' & \leqslant \frac{1}{2}, \label{ApLocB2} \\
-\frac{3}{2} \leqslant & \Delta & \leqslant \frac{1}{2}, \label{ApLocB3} \\
& \delta & \leqslant \frac{1}{2}. \label{ApLocB4}
\end{eqnarray}

Quantum theory's predictions for ideal EPRB experiments satisfy the apparent locality conditions, (\ref{ApLoc1})-(\ref{ApLoc4}), and hence, the bounds predicted by quantum theory, (\ref{QTB1})-(\ref{QTB4}) below, are compatible with bounds (\ref{ApLocB1})-(\ref{ApLocB4}).
\begin{eqnarray}
-2 \sqrt{2} \leqslant & S & \leqslant 2 \sqrt{2}, \label{QTB1} \\
-\frac{1 + \sqrt{2}}{2} \leqslant & \Delta' & \leqslant \frac{\sqrt{2} - 1}{2}, \label{QTB2} \\
-\frac{1 + \sqrt{2}}{2} \leqslant & \Delta & \leqslant \frac{\sqrt{2} - 1}{2}, \label{QTB3} \\
& \delta & \leqslant \frac{\sqrt{2}}{4}. \label{QTB4}
\end{eqnarray}
\section{Bell theorems\label{Bell theorems}}
Theorems which, from the hypothesis of local-realism, derive inequalities constraining the possible values of Bell-quantities in ideal EPRB experiments, are known as Bell theorems.

Realism is the hypothesis that reality exists and, always and everywhere, has well defined properties, regardless of whether these are observed or not \cite{CS78}.

Locality is the hypothesis that two physically separated events must be statistically independent unless some form of influence either propagates from one to the other or propagates from a third event to both.

Locality denies the existence of action-at-a-distance and, in EPRB experiments, requires that outcomes in one arm of the experiment be statistically independent of the measurement parameter value chosen in the other arm \cite{CS78}.

Consequently, any statistic performed on an outcome record collected on the $A$ arm, say on the $A$ outcome record, must not depend on whether the measurement parameter on the $B$ arm was set to $b$ or to $b'$ and thus on whether outcomes in that arm were recorded on the $B$ or on the $B'$ records.
This invariance requirement for all possible statistics is much more demanding than the simple equality of marginal probabilities designated as apparent locality in the previous section.

If the joint probability distributions between $A$ and $B$ and between $A$ and $B'$ were independent, as the joint distribution structure in Table~\ref{JProbTables} allows, such invariance would not be ensured and, accordingly, EPRB experiments may potentially display non-local behavior.

To enforce locality from the point of view of the $A$ arm of the experiment, a realistic model of EPRB experiments must require that outcome records $A$, $B$ and $B'$ be jointly distributed, even though whenever an outcome is recorded on the $B$ record this excludes the possibility of recording any outcome on the $B'$ record.
This requirement of local-realism is known as counterfactual definiteness \cite{Stapp71}.

Enforcing locality from the points of view of both arms of the experiment thus requires that all four outcome records, $A$, $A'$, $B$ and $B'$ be jointly distributed and, in consequence, any local-realistic model of an ideal EPRB experiment must necessarily assume the form of the joint probability table in Table~\ref{LRJProbTable}.

\begin{table}[ht]
\caption{Local-realistic model of an ideal EPRB experiment.\label{LRJProbTable}}
\renewcommand*{\arraystretch}{1.5}
\begin{tabular}{|cc|cccc|c|}
\hline
\multicolumn{2}{|c|}{$p^{AA'BB'}$} & $+$ & $+$ & $-$ & $-$ & $B$ \\
& & $+$ & $-$ & $+$ & $-$ & $B'$ \\
\hline
$+$ & $+$ & $p^{AA'BB'}_{++++}$ & $p^{AA'BB'}_{+++-}$ & $p^{AA'BB'}_{++-+}$ & $p^{AA'BB'}_{++--}$ & $p^{AA'BB'}_{++\pm\pm}$ \\
$+$ & $-$ & $p^{AA'BB'}_{+-++}$ & $p^{AA'BB'}_{+-+-}$ & $p^{AA'BB'}_{+--+}$ & $p^{AA'BB'}_{+---}$ & $p^{AA'BB'}_{+-\pm\pm}$ \\
$-$ & $+$ & $p^{AA'BB'}_{-+++}$ & $p^{AA'BB'}_{-++-}$ & $p^{AA'BB'}_{-+-+}$ & $p^{AA'BB'}_{-+--}$ & $p^{AA'BB'}_{-+\pm\pm}$ \\
$-$ & $-$ & $p^{AA'BB'}_{--++}$ & $p^{AA'BB'}_{--+-}$ & $p^{AA'BB'}_{---+}$ & $p^{AA'BB'}_{----}$ & $p^{AA'BB'}_{--\pm\pm}$ \\
\hline
$A$ & $A'$ & $p^{AA'BB'}_{\pm\pm++}$ & $p^{AA'BB'}_{\pm\pm+-}$ & $p^{AA'BB'}_{\pm\pm-+}$ & $p^{AA'BB'}_{\pm\pm--}$ & $1$ \\
\hline
\end{tabular}
\end{table}

The joint distributions in Table~\ref{JProbTables} which such model predicts, result directly from this joint distribution:
\begin{equation}
\forall_{i,j \in \{+,-\}} \colon
\left\{
\renewcommand*{\arraystretch}{1.5}
\begin{array}{ccl}
p^{AB}_{ij} & = & p^{AA'BB'}_{i\pm j\pm} \\
p^{AB'}_{ij} & = & p^{AA'BB'}_{i\pm\pm j} \\
p^{A'B}_{ij} & = & p^{AA'BB'}_{\pm i j\pm} \\
p^{A'B'}_{ij} & = & p^{AA'BB'}_{\pm\ i\pm j}
\end{array}
\right.. \label{T3toT2}
\end{equation}

The marginal probabilities in Table~\ref{JProbTables} which thus result, satisfy the apparent locality conditions, (\ref{ApLoc1})-(\ref{ApLoc4}), which shows that apparent locality is a necessary consequence of locality.

The probabilities in Table~\ref{LRJProbTable} are constrained only by their non-negativity and their total of 1.

Since Bell-quantities are linear on the probabilities in Table~\ref{JProbTables} and, thus, linear on the probabilities in Table~\ref{LRJProbTable}, linear programming can establish the bounds placed on them by these constraints, leading to the well known Bell inequalities for ideal EPRB experiments:
\begin{eqnarray}
-2 \leqslant & S & \leqslant 2, \label{LRB1} \\
-1 \leqslant & \Delta' & \leqslant 0, \label{LRB2} \\
-1 \leqslant & \Delta & \leqslant 0, \label{LRB3} \\
& \delta & \leqslant \frac{1}{4}. \label{LRB4}
\end{eqnarray}

These are strictly enclosed within the bounds (\ref{ApLocB1})-(\ref{ApLocB4}), which shows that apparent locality, even though necessary, is not a sufficient condition for locality.

Since the bounds predicted by quantum theory, (\ref{QTB1})-(\ref{QTB4}), extend beyond these bounds, and these follow directly from local-realism for ideal EPRB experiments, quantum theory and local-realism are incompatible, as originally shown by Bell \cite{Bell64}.

Realized EPRB experiments, however, have not been ideal.
Bell theorems thus need to be generalized to non-ideal conditions before they can be used to interpret experimental results.

Non-ideal EPRB experiments can be classified into three categories, on the basis of how close to ideal their detection processes are:
\begin{enumerate}
\item
Experiments which have inefficient detection on both arms: These are discussed in Sections~\ref{Inefficient detection on both arms} and~\ref{Perfectly correlated counterfactual detection};
\item
Experiments which have nearly ideal detection on one arm but inefficient detection on the other: Section~\ref{Asymmetric detection efficiencies} discusses these;
\item
Experiments which have nearly ideal detection on both arms but do not follow the EPRB experimental protocol described in Section~\ref{Ideal EPRB experiments}, in particular, do not physically separate the two quantum systems: These are addressed in Section~\ref{Measurement crosstalk}.
\end{enumerate}
\section{Inefficient detection on both arms\label{Inefficient detection on both arms}}
With inefficient detection, a third outcome becomes possible on either arm.
For each analyzed quantum system, in addition to detection on the $+$ channel or detection on the $-$ channel, no detection is now possible.
Let's use the symbol $0$ to represent this outcome and the symbol $*$ to represent summation over all three possible outcomes, $+$, $-$ and $0$.

The joint probability distribution of the four detection records, $A$, $A'$, $B$ and $B'$, that generalizes Table~\ref{LRJProbTable}, now involves $3^4 = 81$ probabilities instead of 16.
For the removable analyzers version, since six detection records are involved, $A$, $A'$, $A''$, $B$, $B'$ and $B''$, $3^4 2^2 = 324$ probabilities are required.

To establish the marginal constraints on these probabilities, additional hypotheses regarding detection are necessary.

If the quantum systems adopted for the experiment do not possess an intrinsic 'detectability' property which might cause some systems to have a higher probability of detection than others, a supplementary hypothesis should be that, in each channel, (i) actual detections are a random sample drawn from the population of all potential detections, with each element having a probability of inclusion $\eta$, the detection efficiency.

The detection efficiencies in each channel, $\eta_{A+}$, $\eta_{A-}$, $\eta_{B+}$ and $\eta_{B-}$, need not be identical.
Nevertheless, to begin, let's make the simplifying assumption of identical detection efficiencies on all channels:
\[\eta_{A+} = \eta_{A-} = \eta_{B+} = \eta_{B-} = \eta.\]

The generalization to different detection efficiencies on either arm of the experiment is made in Section~\ref{Asymmetric detection efficiencies}.

The final additional hypothesis is that (ii) all detectors are independent, in the sense that they neither influence each other nor are influenced by the analyzers or the choice of measurement parameter values in any way other than by the reception of the quantum measurement signals.
Should we say that a detection event has occurred in a detection record when either a $+$ or a $-$ outcome has been recorded in that record, by opposition to a no detection, $0$, outcome, this hypothesis makes all detection vs. no detection events in different detection records, statistically independent.

The conjunction of hypotheses (i) and (ii) is known as fair-sampling detection.

Assuming fair-sampling detection, quantum theory predicts the following bounds on Bell-quantities:
\begin{eqnarray}
-2 \sqrt{2} \, \eta^2 \leqslant & S & \leqslant 2 \sqrt{2} \, \eta^2, \label{QTFSB1} \\
-\frac{\sqrt{2} - 1}{2} \, \eta^2 - \eta \leqslant & \Delta' & \leqslant \frac{1 + \sqrt{2}}{2} \, \eta^2 - \eta, \label{QTFSB2} \\
-\frac{1 + \sqrt{2}}{2} \, \eta^2 \leqslant & \Delta & \leqslant \frac{\sqrt{2} - 1}{2} \, \eta^2, \label{QTFSB3} \\
& \delta & \leqslant \frac{\sqrt{2}}{4} \, \eta^2. \label{QTFSB4}
\end{eqnarray}

Since all these bounds tend to $0$ when detection efficiency decreases, it is useful to introduce three normalized Bell-quantities for which quantum theory predicts bounds independent of detection efficiency.

Because the bounds on $S$, $\Delta$ and $\delta$ are all proportional to $\eta^2$, quantum theory predicts bounds independent of detection efficiency for the following normalized quantities:
\begin{equation}
S_{\text{N}} := \frac{S}{\eta^2}, \qquad \Delta_{\text{N}} := \frac{\Delta}{\eta^2}, \qquad \delta_{\text{N}} := \frac{\delta}{\eta^2}. \label{NDef}
\end{equation}

It is these normalized quantities that have been measured in most Bell tests and the bounds predicted for them by quantum theory are, regardless of detection efficiency:
\begin{eqnarray}
-2 \sqrt{2} \leqslant & S_{\text{N}} & \leqslant 2 \sqrt{2}, \label{QTFSB5} \\
-\frac{1 + \sqrt{2}}{2} \leqslant & \Delta_{\text{N}} & \leqslant \frac{\sqrt{2} - 1}{2}, \label{QTFSB6} \\
& \delta_{\text{N}} & \leqslant \frac{\sqrt{2}}{4}. \label{QTFSB7}
\end{eqnarray}

As for ideal experiments, linear programming can be used to derive the bounds imposed by local-realism and fair-sampling detection on Bell-quantities.

The marginal constraints on the joint probability distribution for the fixed analyzers version become:
\begin{eqnarray}
p^{AA'BB'}_{****} & = & 1, \label{MCFA1} \\
p^{AA'BB'}_{\pm***} = p^{AA'BB'}_{*\pm**} & = & \eta, \label{MCFA2} \\
p^{AA'BB'}_{**\pm*} = p^{AA'BB'}_{***\pm} & = & \eta. \label{MCFA3}
\end{eqnarray}

For the removable analyzers version:
\begin{eqnarray}
p^{AA'A''BB'B''}_{******} & = & 1, \label{MCRA1} \\
p^{AA'A''BB'B''}_{\pm*****} = p^{AA'A''BB'B''}_{*\pm****} & = & \eta, \label{MCRA2} \\
p^{AA'A''BB'B''}_{***\pm**} = p^{AA'A''BB'B''}_{****\pm*} & = & \eta, \label{MCRA3} \\
p^{AA'A''BB'B''}_{**+***} = p^{AA'A''BB'B''}_{*****+} & = & \eta. \label{MCRA4}
\end{eqnarray}

The Bell inequalities implied by these sets of constraints can be derived by solving the respective parametric linear programming problems, and the joint probability solutions that are found to lie on the maximal contour of $S$, involve the following probabilities for joint detection on both arms of the experiment:
\begin{eqnarray*}
p^{AB}_{\pm\pm} = p^{AB'}_{\pm\pm} = p^{A'B}_{\pm\pm} = p^{AB'}_{\pm\pm} & = & \left\{
\renewcommand*{\arraystretch}{1.5}
\begin{array}{lr}
\frac{2}{3} \eta & \colon 0 \leqslant \eta \leqslant \frac{3}{4} \\
2 \eta - 1 & \colon \frac{3}{4} \leqslant \eta \leqslant 1
\end{array} \right. \nonumber \\
&&\neq \eta^2.
\end{eqnarray*}

This means that whether detection occurs or not on the $A$ arm is not yet statistically independent of whether it occurs or not on the $B$ arm.
The formulation developed thus far, enforces locality regarding parameter values, through the structure of the joint probability distribution, but does not yet enforce locality on detection and still allows potentially non-local models.

Adding the following constraints to, respectively, constraints (\ref{MCFA1})-(\ref{MCFA3}) and (\ref{MCRA1})-(\ref{MCRA4}) enforces statistical independence between factual detection on both arms of the experiment:
\begin{equation}
p^{AA'BB'}_{\pm*\pm*} = p^{AA'BB'}_{\pm**\pm} = p^{AA'BB'}_{*\pm\pm*} = p^{AA'BB'}_{*\pm*\pm} = \eta^2; \label{FDFA}
\end{equation}
\begin{eqnarray}
p^{AA'A''BB'B''}_{\pm**\pm**} = p^{AA'A''BB'B''}_{\pm***\pm*} & = & p^{AA'A''BB'B''}_{\pm****+} = \eta^2, \nonumber \\
&& \label{FDRA1} \\
p^{AA'A''BB'B''}_{*\pm*\pm**} = p^{AA'A''BB'B''}_{*\pm**\pm*} & = & p^{AA'A''BB'B''}_{*\pm***+} = \eta^2, \nonumber \\
&& \label{FDRA2} \\
p^{AA'A''BB'B''}_{**+\pm**} = p^{AA'A''BB'B''}_{**+*\pm*} & = & p^{AA'A''BB'B''}_{**+**+} = \eta^2. \nonumber \\
&& \label{FDRA3}
\end{eqnarray}

From these enlarged constraint sets, another set of Bell inequalities results, which includes the well known inequality of Garg and Mermin \cite{GM87}:
\[\left.
\renewcommand*{\arraystretch}{1.5}
\begin{array}{lr}
0 \leqslant \eta \leqslant \frac{2}{3} \colon & -4 \\
\frac{2}{3} \leqslant \eta \leqslant 1 \colon & -\frac{4}{\eta} + 2
\end{array}
\right\}
\leqslant S_{\text{N}} \leqslant
\left\{
\renewcommand*{\arraystretch}{1.5}
\begin{array}{lr}
4 & \colon 0 \leqslant \eta \leqslant \frac{2}{3} \\
\frac{4}{\eta} - 2 & \colon \frac{2}{3} \leqslant \eta \leqslant 1
\end{array}
\right..
\]

However, along the maximal contour of $S$, we find, for $\frac{2}{3} \leqslant \eta \leqslant 1$:
\begin{eqnarray*}
p^{ABB'}_{\pm\pm\pm} = p^{A'BB'}_{\pm\pm\pm} = p^{AA'B}_{\pm\pm\pm} = p^{AA'B'}_{\pm\pm\pm} & = & 2 \eta^2 - \eta \neq \eta^3, \\
p^{AA'BB'}_{\pm\pm\pm\pm} & = & 3 \eta^2 - 2 \eta \neq \eta^4.
\end{eqnarray*}

The joint probabilities for counterfactual detection on one arm are still not statistically independent of detection on the other arm and consequently some non-locality is still allowed.

Enforcing full statistical independence for factual and counterfactual detection on both sides of the experiment requires the addition of the following constraints to (\ref{MCFA1})-(\ref{MCFA3}) and (\ref{FDFA}), for the fixed analyzers version:
\begin{eqnarray*}
p^{AA'BB'}_{\pm\pm**} = p^{AA'BB'}_{**\pm\pm} & = & \eta^2, \\
p^{AA'BB'}_{\pm*\pm\pm} = p^{AA'BB'}_{*\pm\pm\pm} = p^{AA'BB'}_{\pm\pm\pm*} = p^{AA'BB'}_{\pm\pm*\pm} & = & \eta^3 \\
p^{AA'BB'}_{\pm\pm\pm\pm} & = & \eta^4.
\end{eqnarray*}

And the addition of the following constraints to (\ref{MCRA1})-(\ref{MCRA4}) and (\ref{FDRA1})-(\ref{FDRA3}), for the removable analyzers version:
\begin{eqnarray*}
p^{AA'A''BB'B''}_{\pm\pm****} = p^{AA'A''BB'B''}_{\pm*+***} = p^{AA'A''BB'B''}_{*\pm+***} & = & \eta^2, \\
p^{AA'A''BB'B''}_{***\pm\pm*} = p^{AA'A''BB'B''}_{***\pm*+} = p^{AA'A''BB'B''}_{****\pm+} & = & \eta^2, \\
p^{AA'A''BB'B''}_{\pm\pm+***} = p^{AA'A''BB'B''}_{\pm\pm*\pm**} = p^{AA'A''BB'B''}_{\pm\pm**\pm*} & = & \eta^3, \\
p^{AA'A''BB'B''}_{\pm\pm***+} = p^{AA'A''BB'B''}_{\pm*+\pm**} = p^{AA'A''BB'B''}_{\pm*+*\pm*} & = & \eta^3, \\
p^{AA'A''BB'B''}_{\pm*+**+} = p^{AA'A''BB'B''}_{\pm**\pm\pm*} = p^{AA'A''BB'B''}_{\pm**\pm*+} & = & \eta^3, \\
p^{AA'A''BB'B''}_{\pm***\pm+} = p^{AA'A''BB'B''}_{*\pm+\pm**} = p^{AA'A''BB'B''}_{*\pm+*\pm*} & = & \eta^3, \\
p^{AA'A''BB'B''}_{*\pm+**+} = p^{AA'A''BB'B''}_{*\pm*\pm\pm*} = p^{AA'A''BB'B''}_{*\pm*\pm*+} & = & \eta^3, \\
p^{AA'A''BB'B''}_{*\pm**\pm+} = p^{AA'A''BB'B''}_{**+\pm\pm*} = p^{AA'A''BB'B''}_{**+\pm*+} & = & \eta^3, \\
p^{AA'A''BB'B''}_{**+*\pm+} = p^{AA'A''BB'B''}_{***\pm\pm+} & = & \eta^3, \\
p^{AA'A''BB'B''}_{\pm\pm+\pm**} = p^{AA'A''BB'B''}_{\pm\pm+*\pm*} = p^{AA'A''BB'B''}_{\pm\pm+**+} & = & \eta^4, \\
p^{AA'A''BB'B''}_{\pm\pm*\pm\pm*} = p^{AA'A''BB'B''}_{\pm\pm*\pm*+} = p^{AA'A''BB'B''}_{\pm\pm**\pm+} & = & \eta^4, \\
p^{AA'A''BB'B''}_{\pm*+\pm\pm*} = p^{AA'A''BB'B''}_{\pm*+\pm*+} = p^{AA'A''BB'B''}_{\pm*+*\pm+} & = & \eta^4, \\
p^{AA'A''BB'B''}_{\pm**\pm\pm+} = p^{AA'A''BB'B''}_{*\pm+\pm\pm*} = p^{AA'A''BB'B''}_{*\pm+\pm*+} & = & \eta^4, \\
p^{AA'A''BB'B''}_{*\pm+*\pm+} = p^{AA'A''BB'B''}_{*\pm*\pm\pm+} = p^{AA'A''BB'B''}_{**+\pm\pm+} & = & \eta^4, \\
p^{AA'A''BB'B''}_{\pm\pm+\pm\pm*} = p^{AA'A''BB'B''}_{\pm\pm+\pm*+} = p^{AA'A''BB'B''}_{\pm\pm+*\pm+} & = & \eta^5, \\
p^{AA'A''BB'B''}_{\pm\pm*\pm\pm+} = p^{AA'A''BB'B''}_{\pm*+\pm\pm+} = p^{AA'A''BB'B''}_{*\pm+\pm\pm+} & = & \eta^5, \\
p^{AA'A''BB'B''}_{\pm\pm+\pm\pm+} & = & \eta^6.
\end{eqnarray*}

The Bell inequalities that result from the conjunction of local-realism with fair-sampling detection are:
\begin{eqnarray}
2 \eta^4 - 4 \eta^2 \leqslant & S & \leqslant -2 \eta^4 + 4 \eta^2, \label{LRFSB1} \\
-\eta^4 + 2 \eta^3 - 2 \eta \leqslant & \Delta' & \leqslant 0, \label{LRFSB2} \\
-\eta^6 + 3 \eta^4 - 3 \eta^2 \leqslant & \Delta & \leqslant \eta^6 - 2 \eta^5 + 2 \eta^4 -4 \eta^3 + 3 \eta^2, \nonumber \\
&& \label{LRFSB3} \\
& \delta & \leqslant \frac{\eta^6}{2} - \frac{\eta^5}{2} - \frac{\eta^4}{4} - \eta^3 + \frac{3 \eta^2}{2}, \nonumber \\
&& \label{LRFSB4}
\end{eqnarray}
and, for normalized Bell-quantities:
\begin{eqnarray}
2 \eta^2 - 4 \leqslant & S_{\text{N}} & \leqslant -2 \eta^2 + 4, \label{LRFSB5} \\
-\eta^4 + 3 \eta^2 - 3 \leqslant & \Delta_{\text{N}} & \leqslant \eta^4 - 2 \eta^3 + 2 \eta^2 -4 \eta + 3, \label{LRFSB6} \\
& \delta_{\text{N}} & \leqslant \frac{\eta^4}{2} - \frac{\eta^3}{2} - \frac{\eta^2}{4} - \eta + \frac{3}{2}. \label{LRFSB7}
\end{eqnarray}

\begin{figure}[ht]
\includegraphics[bb = 60 365 535 790, scale = 0.32, clip]{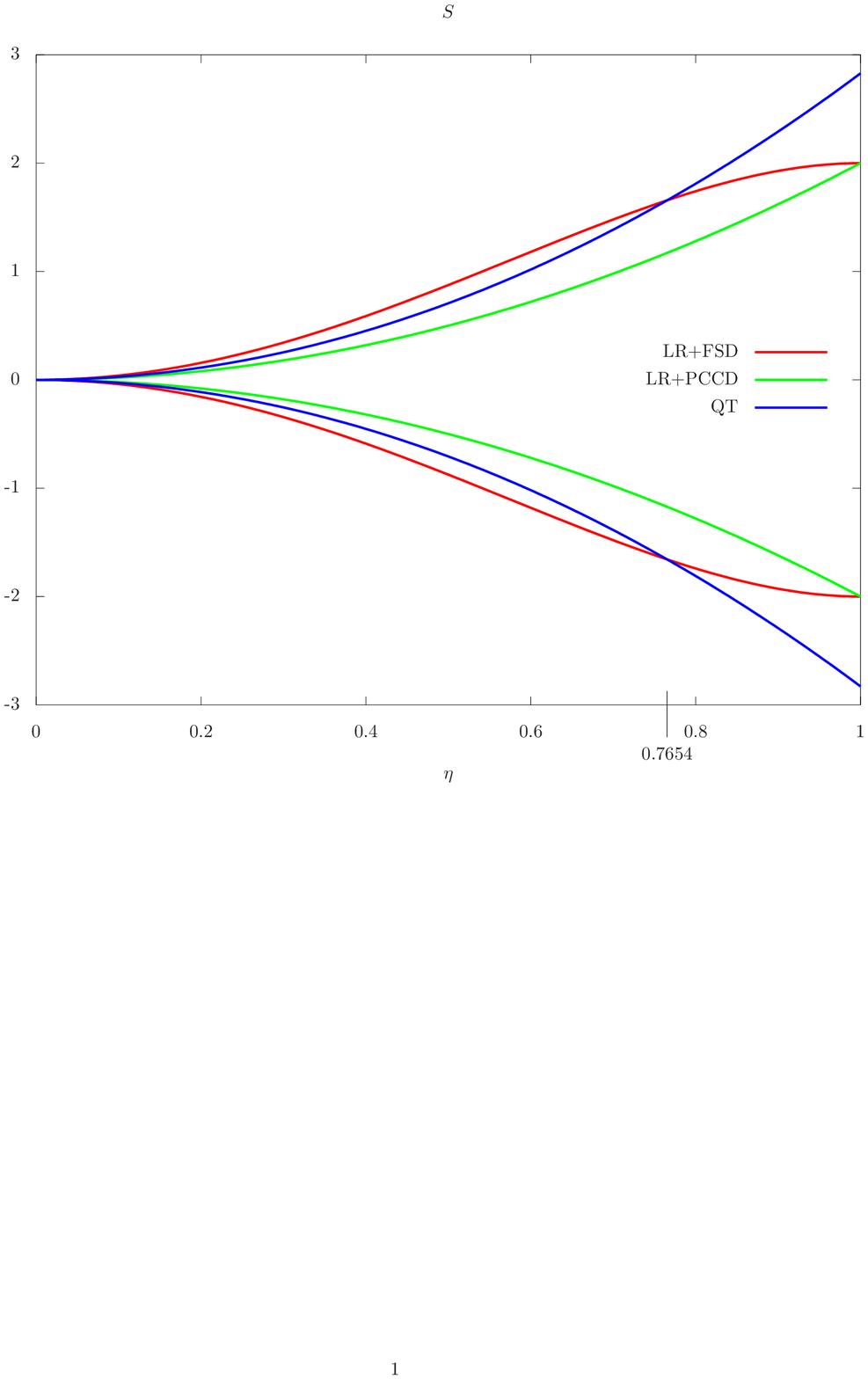}
\caption{Bounds on Bell-quantity $S$.\label{S}}
\end{figure}

\begin{figure}[ht]
\includegraphics[bb = 60 365 535 790, scale = 0.32, clip]{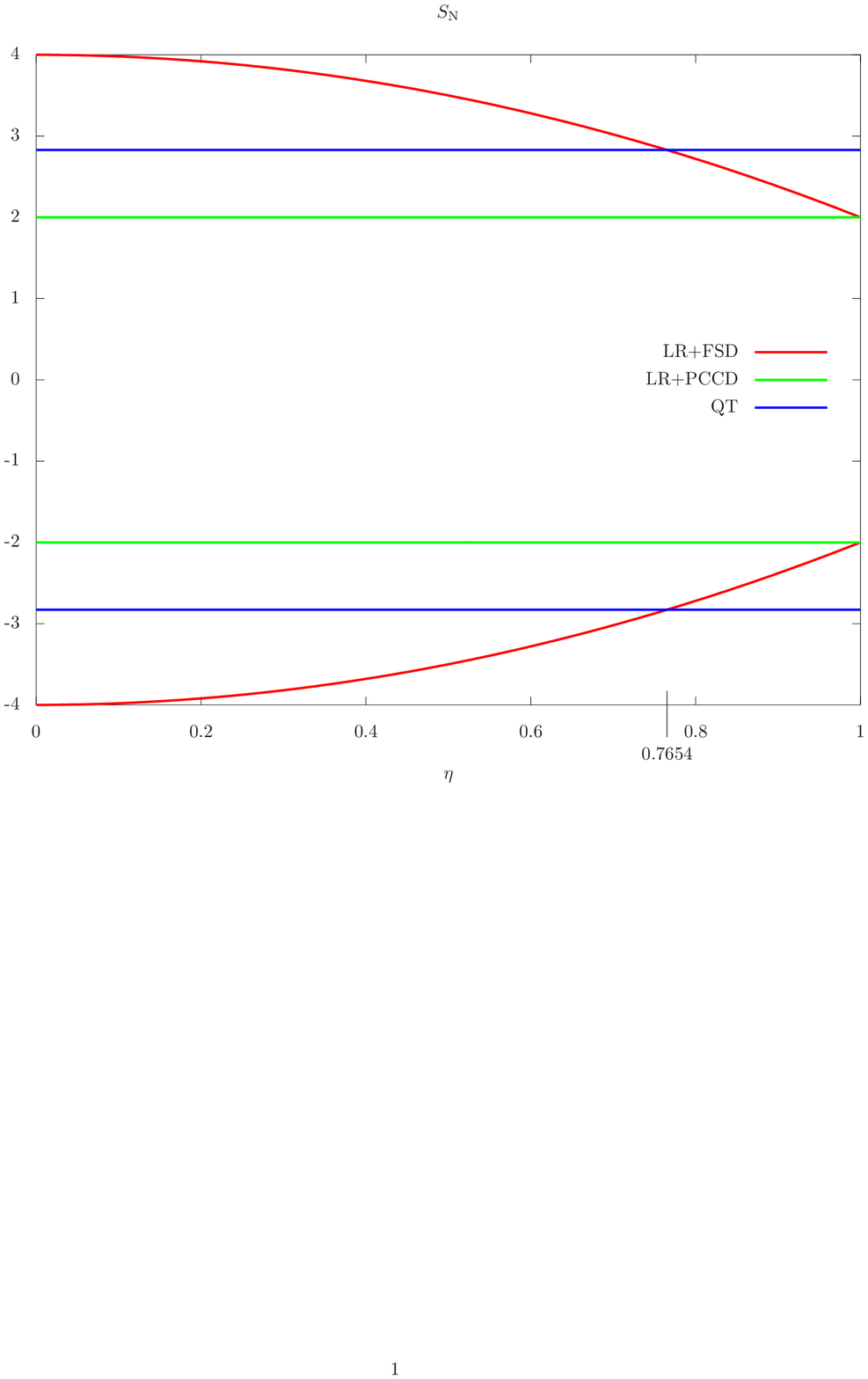}
\caption{Bounds on Bell-quantity $S_{N}$.\label{SN}}
\end{figure}

\begin{figure}[ht]
\includegraphics[bb = 60 365 535 790, scale = 0.32, clip]{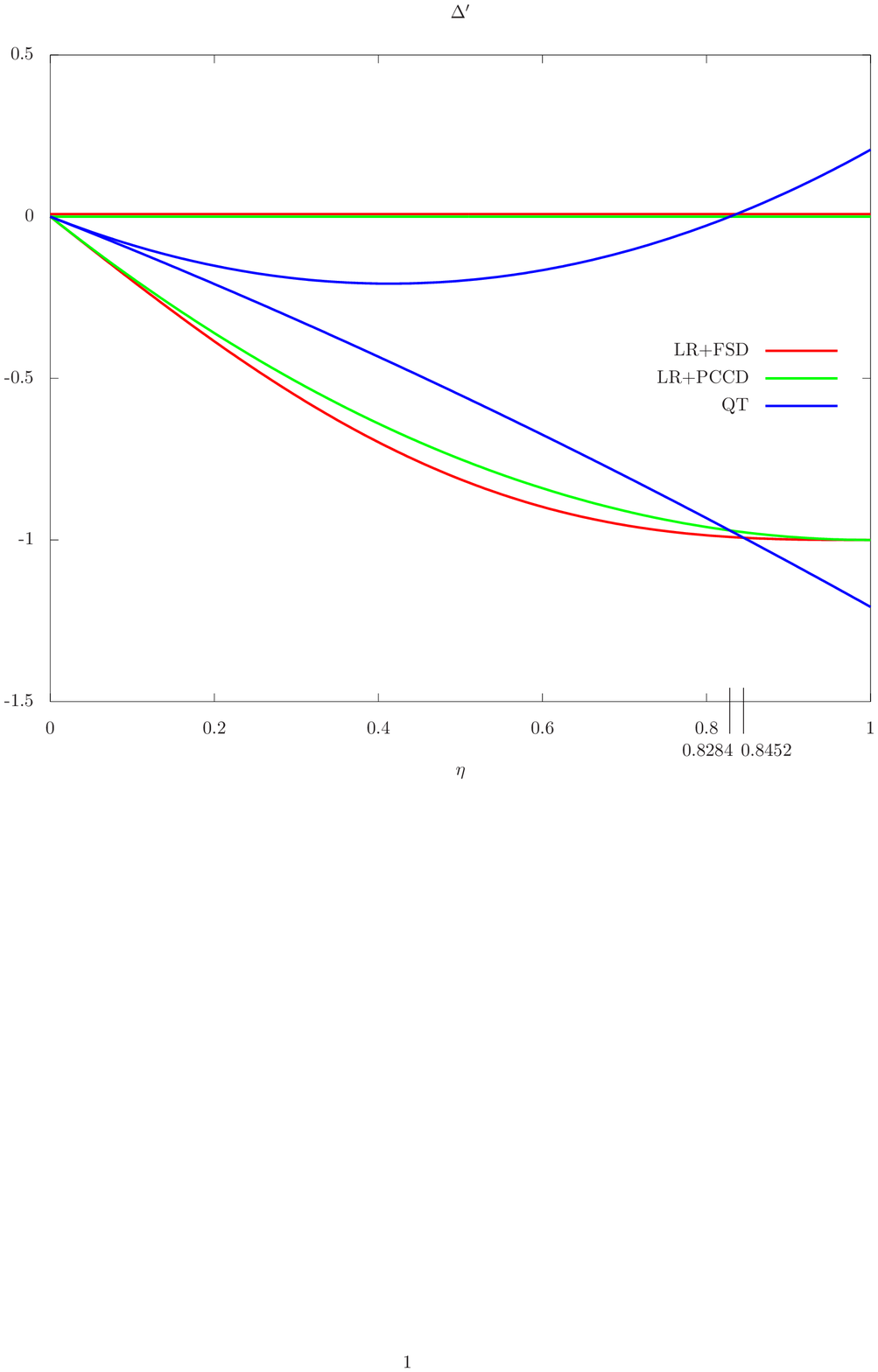}
\caption{Bounds on Bell-quantity $\Delta'$.\label{Delta'}}
\end{figure}

\begin{figure}[ht]
\includegraphics[bb = 60 365 535 790, scale = 0.32, clip]{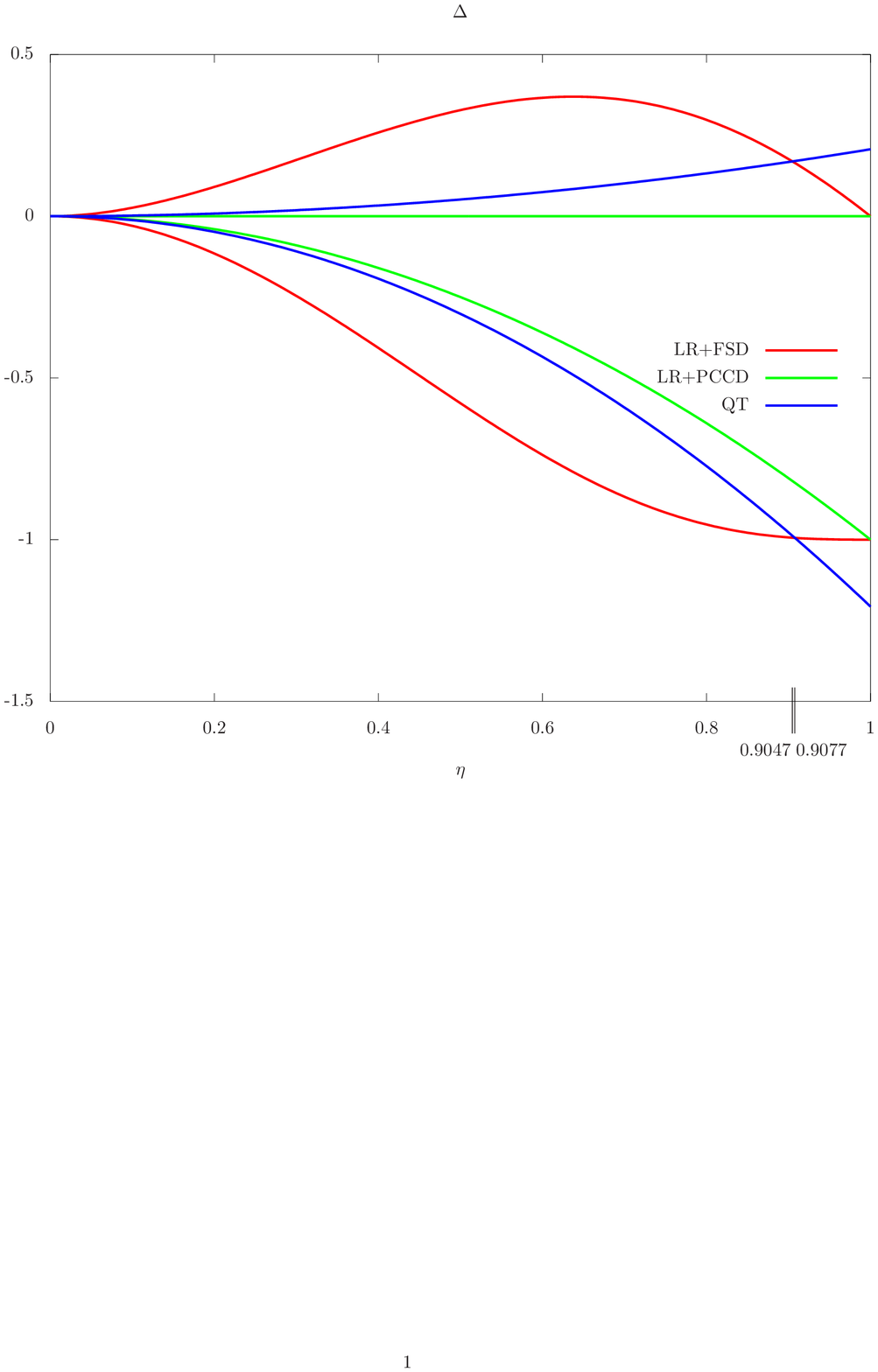}
\caption{Bounds on Bell-quantity $\Delta$.\label{Delta}}
\end{figure}

\begin{figure}[ht]
\includegraphics[bb = 60 365 535 790, scale = 0.32, clip]{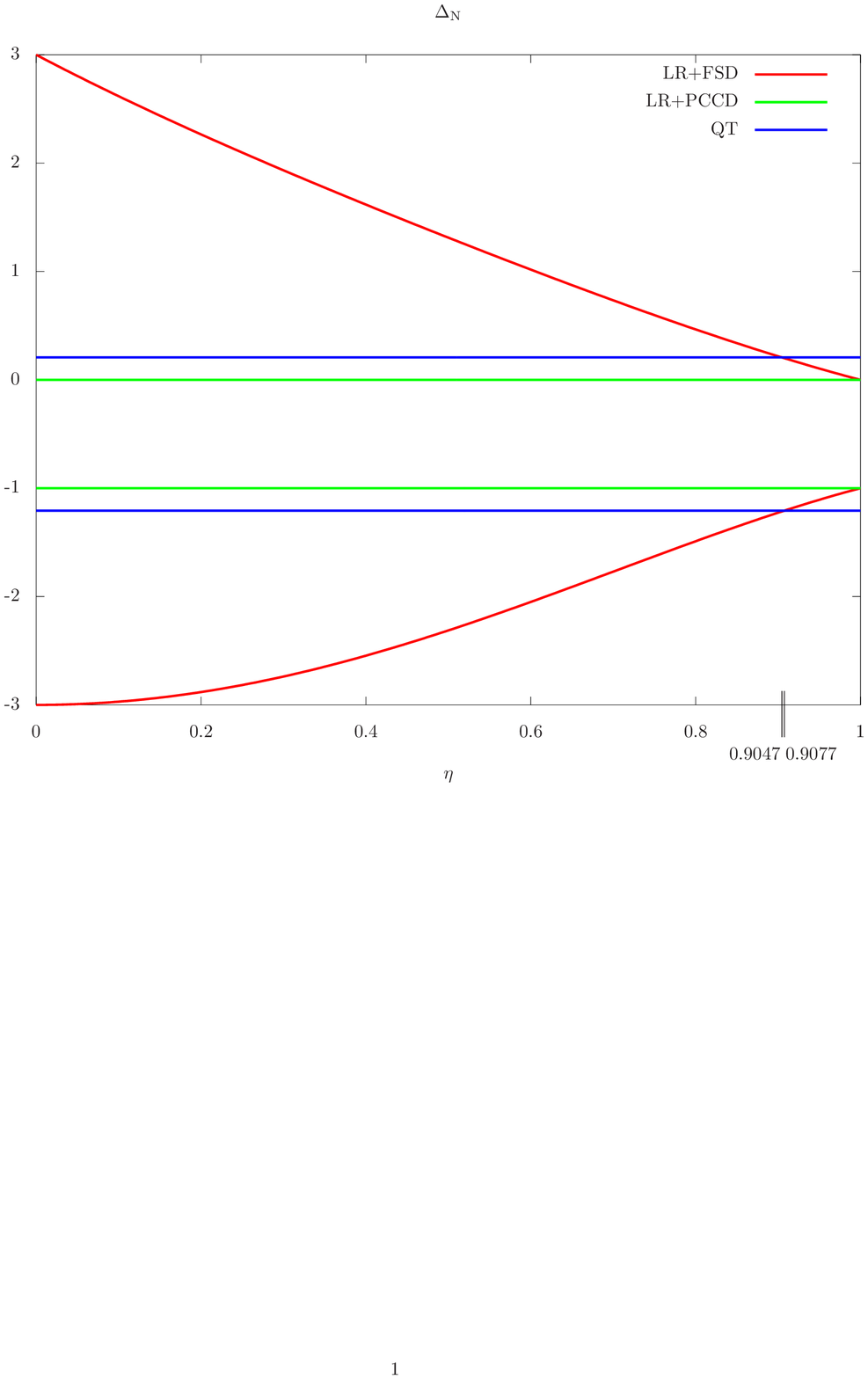}
\caption{Bounds on Bell-quantity $\Delta_{N}$.\label{DeltaN}}
\end{figure}

\begin{figure}[ht]
\includegraphics[bb = 60 365 535 790, scale = 0.32, clip]{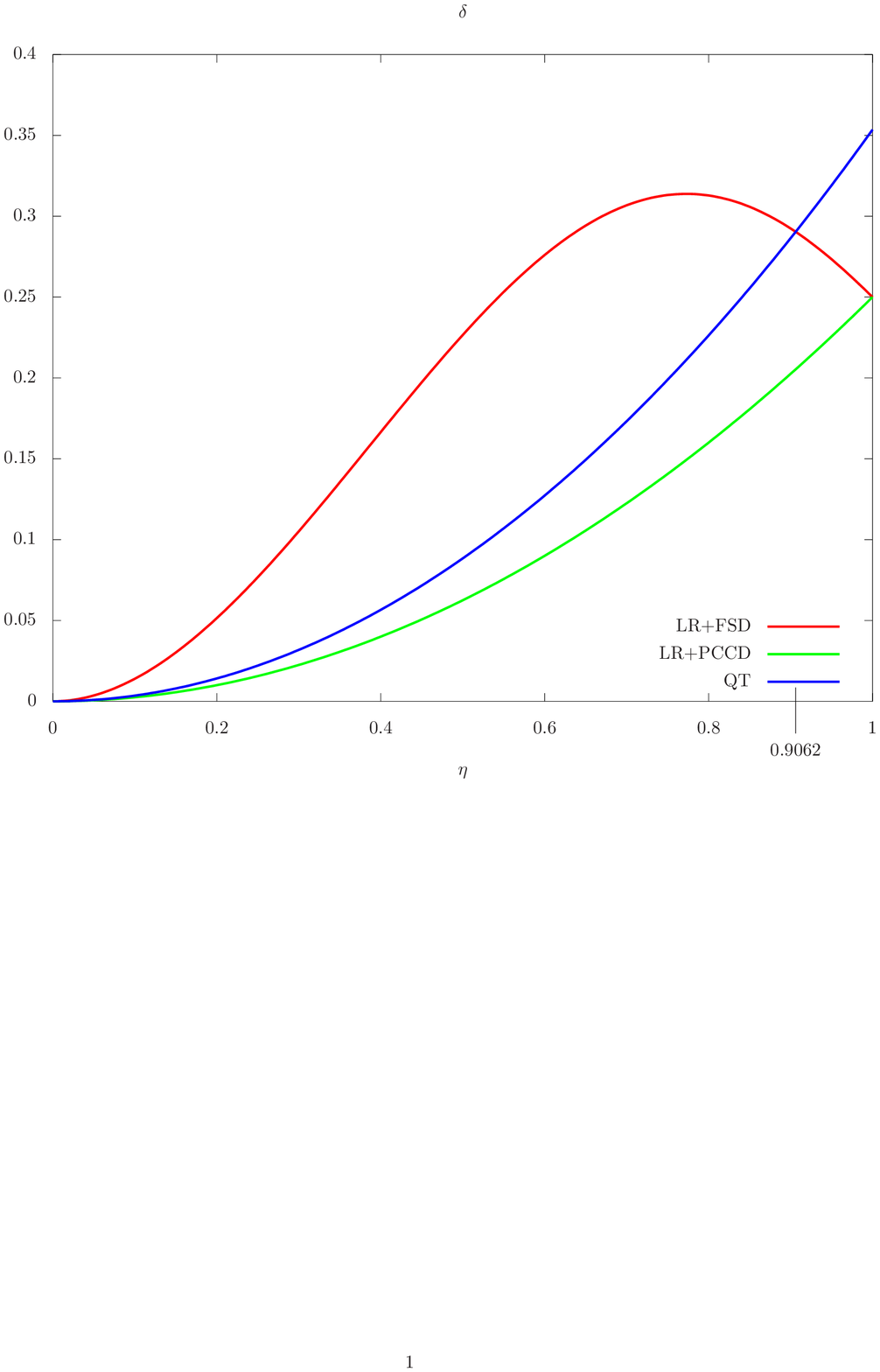}
\caption{Bounds on Bell-quantity $\delta$.\label{delta}}
\end{figure}

\begin{figure}[ht]
\includegraphics[bb = 60 365 535 790, scale = 0.32, clip]{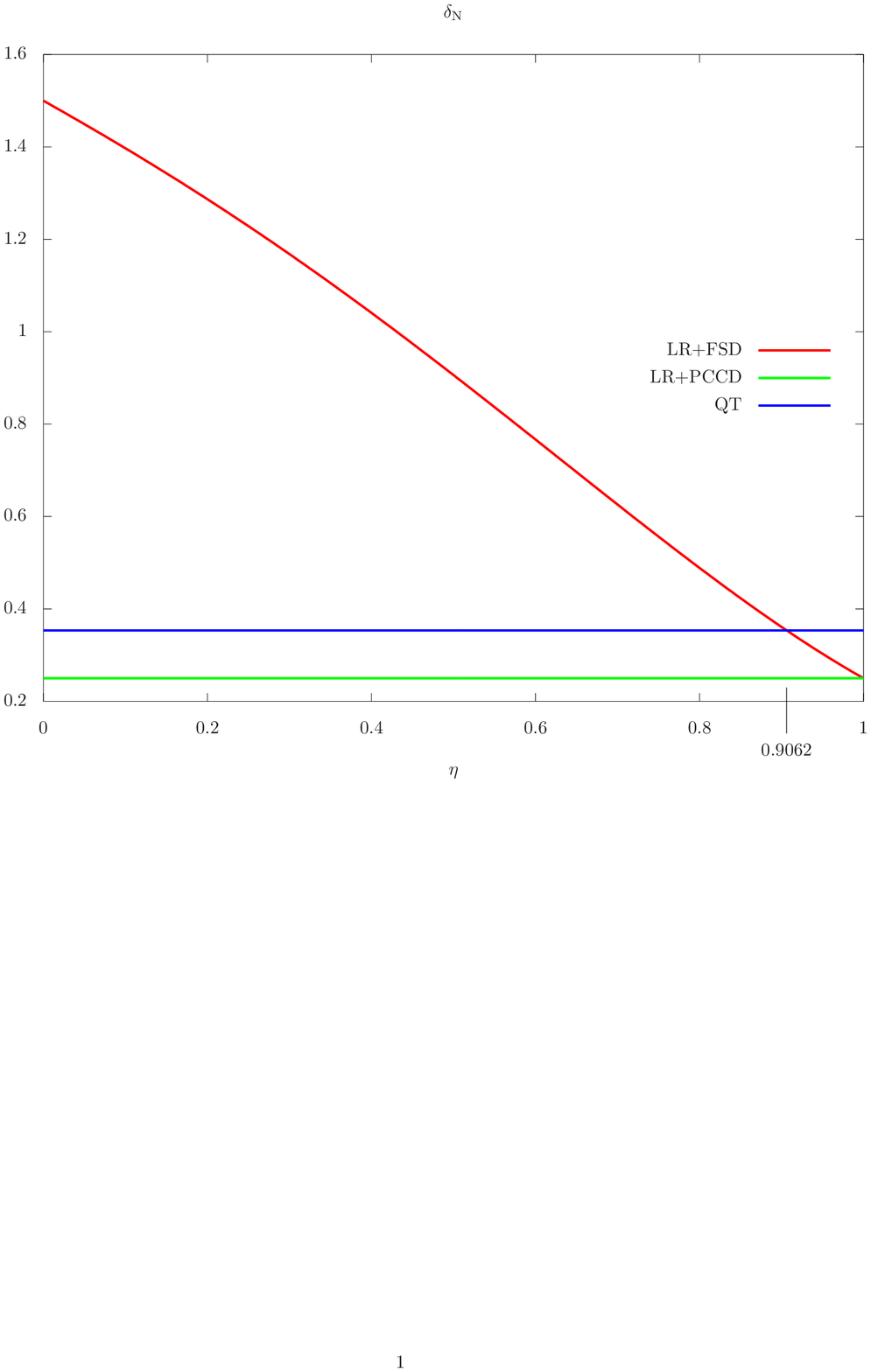}
\caption{Bounds on Bell-quantity $\delta_{N}$.\label{deltaN}}
\end{figure}

The bounds these inequalities impose are graphically depicted, for each Bell-quantity, by the LR+FSD (local-realism and fair-sampling detection) lines in Figs.~\ref{S}~to~\ref{deltaN}.

The respective bounds predicted by quantum theory, (\ref{QTFSB1})-(\ref{QTFSB4}) and (\ref{QTFSB5})-(\ref{QTFSB7}), are also depicted in the same Figures by the QT lines.

The predictions of quantum theory can be seen to be compatible with the bounds imposed by local-realism and fair-sampling detection for all detection efficiencies below a critical threshold, specific to each Bell-quantity.
The values of the critical detection efficiencies, $\eta_{c}$, are listed in Table~\ref{CritDetEffTable}.

\begin{table}[ht]
\caption{Critical detection efficiencies.\label{CritDetEffTable}}
\begin{ruledtabular}
\renewcommand*{\arraystretch}{1.5}
\begin{tabular}{cccc}
\multicolumn{2}{c}{Bell-quantity} & \multicolumn{2}{c}{$\eta_{\text{c}}$} \\
\hline
\multicolumn{2}{c}{$S$, $S_{\text{N}}$} & $\sqrt{2-\sqrt{2}}$ & $\approx 0.7654$ \\
$\Delta'$ & Upper Bnd. & $2(\sqrt{2}-1)$ & $\approx 0,8284$ \\
$\Delta'$ & Lower Bnd. & & $\approx 0.8452$ \\
$\Delta$, $\Delta_{\text{N}}$ & Upper Bnd. & & $\approx 0.9047$ \\
$\Delta$, $\Delta_{\text{N}}$ & Lower Bnd. & & $\approx 0.9077$ \\
\multicolumn{2}{c}{$\delta$, $\delta_{\text{N}}$} & & $\approx 0.9062$ \\
\end{tabular}
\end{ruledtabular}
\end{table}
\clearpage
\begin{description}
\item[Abbreviations]
\item[LR+FSD]
Conjunction of the hypotheses of local-realism and fair-sampling detection;
\item[LR+PCCD]
Conjunction of the hypotheses of local-realism and perfectly correlated counterfactual detection (see Section~\ref{Perfectly correlated counterfactual detection});
\item[QT]
Bounds predicted by quantum theory.
\end{description}
\clearpage
The lowest critical detection efficiency, and hence the most selective Bell test, can be seen to be that based on the two-channel analyzer EPRB experiment version and Bell-quantity $S$ or $S_{\text{N}}$.

Because the bounds imposed on Bell-quantities by the conjunction of local-realism and fair-sampling detection are dependent on detection efficiency, experiments which seek to discriminate between quantum theory and local-realism through the measurement of Bell-quantities in EPRB experiments, known as Bell tests, must also measure detection efficiency.

If detection efficiency is found to be below the above critical thresholds for all detectors involved, consistency of the observed results with the predictions of quantum theory is insufficient to discriminate between quantum theory and the conjunction of local-realism and fair-sampling detection.

Consequently, because no photon detector has yet achieved the required efficiency \cite{KSCEP93}, all purely optical Bell tests performed to this date have been inconclusive due to insufficient detection efficiency.
These include \cite{FC72,C76,FT76,AGR81,AGR82,ADR82,PDBK85,HDPBK87,OM88,SA88,RT90,KMWZSS95,WJSWZ98,TBGZ99,KWWAE99,AJK05,MCBLJKK05,SBHGZ08} among others.

To this category also belong tests in which the state of the quantum systems can be detected with near certainty when they are examined, but they are examined only in a small subset of prepared instances, heralded, usually, by photon detections \cite{CLDCRF07,MMMOM08}.
The low probability for prepared quantum systems to actually be examined is equivalent to low detection efficiency and makes this type of test inconclusive as well.

It is important to note that this result is quite different from what is known as 'the detection loophole' \cite{Pearle70} because the failure to reject local-realism by these tests has here been shown to result from the conjunction of local-realism and fair-sampling detection, whereas it had been understood that unrepresentative sampling detection was required to invalidate these tests.
\section{Perfectly correlated counterfactual detection\label{Perfectly correlated counterfactual detection}}
The result presented in the previous section is somewhat surprising because for several decades it has been consensual that the inequalities applicable to normalized Bell-quantities under the conjunction of local-realism with additional hypotheses similar to fair-sampling detection, such as 'no-enhancement' detection \cite{CH74} and 'microscopical symmetric detection' \cite{GR81}, were:
\begin{eqnarray}
-2 \leqslant & S_{\text{N}} & \leqslant 2, \label{LRPCCDB5} \\
-1 \leqslant & \Delta_{\text{N}} & \leqslant 0, \label{LRPCCDB6} \\
& \delta_{\text{N}} & \leqslant \frac{1}{4}. \label{LRPCCDB7}
\end{eqnarray}

To understand the difference that leads to one set of inequalities instead of the other, let's re-derive the result for the Bell-quantity $S$ obtained in Section~\ref{Inefficient detection on both arms}, (\ref{LRFSB1}), but, this time, using a variation on Mermin's logic ladder \cite{Mermin81} instead of linear programming.

Please consider four binary detection records, $A$, $A'$, $B$ and $B'$, all with the same number of recorded symbols, $N_{\text{P}}$.
Given their binary nature, the number of positions in which $A$ and $B'$ agree or disagree is constrained by the numbers of positions in which $A$ and $B$ agree or disagree, $A'$ and $B$ agree or disagree and $A'$ and $B'$ agree or disagree:
\[\left\{
\renewcommand*{\arraystretch}{1.5}
\begin{array}{rcl}
N^{AB'}_{\text{Agree}} & \leqslant & N^{AB}_{\text{Agree}} + N^{A'B}_{\text{Agree}} + N^{A'B'}_{\text{Agree}} \\
N^{AB'}_{\text{Disagree}} & \leqslant & N^{AB}_{\text{Disagree}} + N^{A'B}_{\text{Disagree}} + N^{A'B'}_{\text{Disagree}}
\end{array}
\right..\]

Dividing by the total number of symbols, $N_{\text{P}}$, and taking the limit $N_{\text{P}} \to \infty$, transforms the above inequalities on counts into inequalities on probabilities:
\[\left\{
\renewcommand*{\arraystretch}{1.5}
\begin{array}{rcl}
p^{AB'}_{\text{Agree}} & \leqslant & p^{AB}_{\text{Agree}} + p^{A'B}_{\text{Agree}} + p^{A'B'}_{\text{Agree}} \\
p^{AB'}_{\text{Disagree}} & \leqslant & p^{AB}_{\text{Disagree}} + p^{A'B}_{\text{Disagree}} + p^{A'B'}_{\text{Disagree}}
\end{array}
\right..\]

If, instead of being binary, the four detection records are ternary but we remain interested only in agreements or disagreements involving two symbols, say $+$ and $-$, then the above inequalities still apply but only to the subset of events in which all four records jointly have only $+$ and $-$ symbols.

For $A$ and $B'$ to be comparable, both have to have either a $+$ or a $-$ symbol, but if, in some of these instances, either $A'$ or $B$ have a 'no detection' outcome, the above inequalities do not, in these specific instances, constrain the agreement or disagreement between $A$ and $B'$.

The probability for $A$ and $B'$ to jointly have detection outcomes, either $+$ or $-$, is $\eta^2$ and the probability for all four detection records to jointly have only detection outcomes is $\eta^4$.

Thus, under inefficient detection, in instances with probability $\eta^2 - \eta^4$, the agreement or disagreement between $A$ and $B'$ is not constrained by the above inequalities and hence this probability must be added to the right-hand sides of both inequalities:
\[\left\{
\renewcommand*{\arraystretch}{1.5}
\begin{array}{rcl}
p^{AB'}_{\text{Agree}} & \leqslant & p^{AB}_{\text{Agree}} + p^{A'B}_{\text{Agree}} + p^{A'B'}_{\text{Agree}} + \eta^2 - \eta^4 \\
p^{AB'}_{\text{Disagree}} & \leqslant & p^{AB}_{\text{Disagree}} + p^{A'B}_{\text{Disagree}} + p^{A'B'}_{\text{Disagree}} \\
&& + \eta^2 - \eta^4
\end{array}
\right..\]

Since,
\[\forall_{X \in \{A,A'\}, Y \in \{B,B'\}} \colon p^{XY}_{\pm\pm} = p^{XY}_{\text{Agree}} + p^{XY}_{\text{Disagree}} = \eta^2,\]
and the correlation between outcome records is:
\[\forall_{X \in \{A,A'\}, Y \in \{B,B'\}} \colon E^{XY} = p^{XY}_{\text{Agree}} - p^{XY}_{\text{Disagree}},\]
solving for $p^{XY}_{\text{Agree}}$ and $p^{XY}_{\text{Disagree}}$ gives:
\[\forall_{X \in \{A,A'\}, Y \in \{B,B'\}} \colon
\left\{
\renewcommand*{\arraystretch}{1.5}
\begin{array}{rcl}
p^{XY}_{\text{Agree}} = \frac{\eta^2 + E^{XY}}{2} \\
p^{XY}_{\text{Disagree}} = \frac{\eta^2 - E^{XY}}{2}
\end{array}
\right..\]

Substituting into the inequalities results in:
\[\left\{
\renewcommand*{\arraystretch}{1.5}
\begin{array}{rcl}
\frac{\eta^2 + E^{AB'}}{2} & \leqslant & \frac{\eta^2 + E^{AB}}{2} + \frac{\eta^2 + E^{A'B}}{2} + \frac{\eta^2 + E^{A'B'}}{2} \\
&& + \eta^2 - \eta^4 \\
\frac{\eta^2 - E^{AB'}}{2} & \leqslant & \frac{\eta^2 - E^{AB}}{2} + \frac{\eta^2 - E^{A'B}}{2} + \frac{\eta^2 - E^{A'B'}}{2} \\
&& + \eta^2 - \eta^4
\end{array}
\right..\]

Collecting all correlations on the left-hand sides and recognizing the definition of $S$, (\ref{SDef}), results in inequality~(\ref{LRFSB1}) from Section~\ref{Inefficient detection on both arms}:
\[\left\{
\renewcommand*{\arraystretch}{1.5}
\begin{array}{rcl}
-S & \leqslant & 4 \eta^2 - 2 \eta^4 \\
S & \leqslant & 4 \eta^2 - 2 \eta^4
\end{array}
\right.,\]

However, had we not added the terms $\eta^2 - \eta^4$ on the right-hand sides of the inequalities, the result would have been:
\[\left\{
\renewcommand*{\arraystretch}{1.5}
\begin{array}{rcl}
-S & \leqslant & 2 \eta^2\\
S & \leqslant & 2 \eta^2
\end{array}
\right.,\]
which, divided by $\eta^2$, produces inequality~(\ref{LRPCCDB5}) on the normalized quantity $S_{\text{N}}$.

It is thus the presence or absence of the $\eta^2 - \eta^4$ terms that distinguishes one set of inequalities from the other.

The origin of these terms lies in the joint probability for $A$ and $B'$ both having detection outcomes being $\eta^2$ and the joint probability for all four detection outcome records $A$, $A'$, $B$ and $B'$ to all have detection outcomes being $\eta^4$.
This is a direct consequence of the fair-sampling detection hypothesis.

However, let's consider the alternative hypothesis that every time $A$ has a detection outcome $A'$ would also have had a detection outcome had the measurement parameter been set to $a'$ instead of $a$.

Because, from a realistic point of view, an outcome that did not take place cannot possibly influence one that did, such perfect correlation between counterfactual detections can only come from the quantum systems having a 'detectability' property which fully determines whether each system will be detected or not.

This hypothesis contradicts fair-sampling and is consequently a form of unrepresentative sampling detection.
It would also make detection efficiency a property of the source of quantum systems instead of a property of the detectors, in contradiction to all known experimental evidence regarding particle detection.

In EPRB experiments, from a local-realistic point of view, the source would have to produce pairs of quantum systems totally uncorrelated in this 'detectability' property while totally correlated in the entangled property and measurement would have to be totally random while detection would have to be completely deterministic.

This combination, even if nearly self-contradictory, is admissible under local-realism but by no means required.

Let's designate by 'perfectly correlated counterfactual detection' the hypothesis that detection in $A$ would always have meant detection in $A'$ and detection in $B'$ would also always have meant detection in $B$.

The conjunction of local-realism with perfectly correlated counterfactual detection implies the following constraints on local-realistic models of the fixed analyzers EPRB experiment version:
\begin{eqnarray*}
p^{AA'BB'}_{****} & = & 1, \\
p^{AA'BB'}_{\pm***} = p^{AA'BB'}_{*\pm**} = p^{AA'BB'}_{**\pm*} = p^{AA'BB'}_{***\pm} & = & \eta, \\
p^{AA'BB'}_{\pm*\pm*} = p^{AA'BB'}_{\pm**\pm} = p^{AA'BB'}_{*\pm\pm*} = p^{AA'BB'}_{*\pm*\pm} & = & \eta^2, \\
p^{AA'BB'}_{\pm\pm**} = p^{AA'BB'}_{**\pm\pm} & = & \eta, \\
p^{AA'BB'}_{\pm*\pm\pm} = p^{AA'BB'}_{*\pm\pm\pm} = p^{AA'BB'}_{\pm\pm\pm*} = p^{AA'BB'}_{\pm\pm*\pm} & = & \eta^2, \\
p^{AA'BB'}_{\pm\pm\pm\pm} & = & \eta^2.
\end{eqnarray*}

For the removable analyzers version, the constraints are:
\begin{eqnarray*}
p^{AA'A''BB'B''}_{******} & = & 1, \\
p^{AA'A''BB'B''}_{\pm*****} = p^{AA'A''BB'B''}_{*\pm****} = p^{AA'A''BB'B''}_{**+***} & = & \eta, \\ 
p^{AA'A''BB'B''}_{***\pm**} = p^{AA'A''BB'B''}_{****\pm*} = p^{AA'A''BB'B''}_{*****+} & = & \eta, \\
p^{AA'A''BB'B''}_{\pm**\pm**} = p^{AA'A''BB'B''}_{\pm***\pm*} = p^{AA'A''BB'B''}_{\pm****+} & = & \eta^2, \\
p^{AA'A''BB'B''}_{*\pm*\pm**} = p^{AA'A''BB'B''}_{*\pm**\pm*} = p^{AA'A''BB'B''}_{*\pm***+} & = & \eta^2, \\
p^{AA'A''BB'B''}_{**+\pm**} = p^{AA'A''BB'B''}_{**+*\pm*} = p^{AA'A''BB'B''}_{**+**+} & = & \eta^2, \\
p^{AA'A''BB'B''}_{\pm\pm****} = p^{AA'A''BB'B''}_{\pm*+***} = p^{AA'A''BB'B''}_{*\pm+***} & = & \eta, \\
p^{AA'A''BB'B''}_{***\pm\pm*} = p^{AA'A''BB'B''}_{***\pm*+} = p^{AA'A''BB'B''}_{****\pm+} & = & \eta, \\
p^{AA'A''BB'B''}_{\pm\pm*\pm**} = p^{AA'A''BB'B''}_{\pm\pm**\pm*} = p^{AA'A''BB'B''}_{\pm\pm***+} & = & \eta^2, \\
p^{AA'A''BB'B''}_{\pm*+\pm**} = p^{AA'A''BB'B''}_{\pm*+*\pm*} = p^{AA'A''BB'B''}_{\pm*+**+} & = & \eta^2, \\
p^{AA'A''BB'B''}_{\pm**\pm\pm*} = p^{AA'A''BB'B''}_{\pm**\pm*+} = p^{AA'A''BB'B''}_{\pm***\pm+} & = & \eta^2, \\
p^{AA'A''BB'B''}_{*\pm+\pm**} = p^{AA'A''BB'B''}_{*\pm+*\pm*} = p^{AA'A''BB'B''}_{*\pm+**+} & = & \eta^2, \\
p^{AA'A''BB'B''}_{*\pm*\pm\pm*} = p^{AA'A''BB'B''}_{*\pm*\pm*+} = p^{AA'A''BB'B''}_{*\pm**\pm+} & = & \eta^2, \\
p^{AA'A''BB'B''}_{**+\pm\pm*} = p^{AA'A''BB'B''}_{**+\pm*+} = p^{AA'A''BB'B''}_{**+*\pm+} & = & \eta^2, \\
p^{AA'A''BB'B''}_{\pm\pm+***} = p^{AA'A''BB'B''}_{***\pm\pm+} & = & \eta, \\
p^{AA'A''BB'B''}_{\pm\pm+\pm**} = p^{AA'A''BB'B''}_{\pm\pm+*\pm*} = p^{AA'A''BB'B''}_{\pm\pm+**+} & = & \eta^2, \\
p^{AA'A''BB'B''}_{\pm\pm*\pm\pm*} = p^{AA'A''BB'B''}_{\pm\pm*\pm*+} = p^{AA'A''BB'B''}_{\pm\pm**\pm+} & = & \eta^2, \\
p^{AA'A''BB'B''}_{\pm*+\pm\pm*} = p^{AA'A''BB'B''}_{\pm*+\pm*+} = p^{AA'A''BB'B''}_{\pm*+*\pm+} & = & \eta^2, \\
p^{AA'A''BB'B''}_{\pm**\pm\pm+} = p^{AA'A''BB'B''}_{*\pm+\pm\pm*} = p^{AA'A''BB'B''}_{*\pm+\pm*+} & = & \eta^2, \\
p^{AA'A''BB'B''}_{*\pm+*\pm+} = p^{AA'A''BB'B''}_{*\pm*\pm\pm+} = p^{AA'A''BB'B''}_{**+\pm\pm+} & = & \eta^2, \\
p^{AA'A''BB'B''}_{\pm\pm+\pm\pm*} = p^{AA'A''BB'B''}_{\pm\pm+\pm*+} = p^{AA'A''BB'B''}_{\pm\pm+*\pm+} & = & \eta^2, \\
p^{AA'A''BB'B''}_{\pm\pm*\pm\pm+} = p^{AA'A''BB'B''}_{\pm*+\pm\pm+} = p^{AA'A''BB'B''}_{*\pm+\pm\pm+} & = & \eta^2, \\
p^{AA'A''BB'B''}_{\pm\pm+\pm\pm+} & = & \eta^2.
\end{eqnarray*}

The Bell inequalities implied by the conjunction of local-realism and perfectly correlated counterfactual detection are:
\begin{eqnarray*}
-2 \eta^2 \leqslant & S & \leqslant 2 \eta^2, \\*
\eta^2 - 2 \eta \leqslant & \Delta' & \leqslant 0, \\*
-\eta^2 \leqslant & \Delta & \leqslant 0, \\*
& \delta & \leqslant \frac{\eta^2}{4}.
\end{eqnarray*}

From these follow inequalities (\ref{LRPCCDB5})-(\ref{LRPCCDB7}), for the normalized quantities, confirming that these inequalities are indeed implied by the conjunction of local-realism with perfectly correlated counterfactual detection.

These bounds are graphically depicted by the LR+PCCD (local-realism and perfectly correlated counterfactual detection) lines in Figures~\ref{S}~to~\ref{deltaN}.

As may be seen, the bounds predicted by quantum theory (QT lines) allow violation of these bounds for all detection efficiency values.

In fact, optical Bell tests \cite{FC72,C76,FT76,AGR81,AGR82,ADR82,PDBK85,HDPBK87,OM88,SA88,RT90,KMWZSS95,WJSWZ98,TBGZ99,KWWAE99,AJK05,MCBLJKK05,SBHGZ08} have systematically violated inequalities (\ref{LRPCCDB5})-(\ref{LRPCCDB7}) and it is, thus, a well established fact that the conjunction of local-realism and perfectly correlated counterfactual detection has been empirically rejected.

Since these optical Bell tests were shown, in Section~\ref{Inefficient detection on both arms}, to be compatible with the conjunction of local-realism and fair-sampling detection, the above fact simply implies empirical rejection of perfectly correlated counterfactual detection, which is not surprising given the nearly self-contradictory nature of this hypothesis.

The confusion between the hypotheses of fair-sampling detection and perfectly correlated counterfactual detection was due to neither \cite{CH74} nor \cite{GR81} having explicitly addressed the issue of counterfactuality.
It thus ended entering these works as an implicit assumption of perfectly correlated counterfactual detection whereas the authors had clearly intended to introduce a fair-sampling detection hypothesis instead.
Given the strong empirical evidence available in favor of fair-sampling detection, this confusion directly led to the incorrect interpretation of experimental evidence as empirical rejection of local-realism.

In the end, it is the claims of rejection of local-realism by these Bell tests which, in fact, require unrepresentative sampling detection.
\section{Asymmetric detection efficiencies\label{Asymmetric detection efficiencies}}
In Section~\ref{Inefficient detection on both arms}, the simplifying assumption of identical detection efficiencies on all channels was made.

Another important generalization of Bell theorems is, however, the treatment of asymmetric EPRB experiments, in which detection is nearly ideal on one arm while inefficient on the other.

To address this scenario and clearly distinguish Bell tests in this category from those discussed in Section~\ref{Inefficient detection on both arms}, let's now consider the general case of different detection efficiencies on either arm of the EPRB experiment:
\[\eta_{A+} = \eta_{A-} = \eta_A, \qquad \eta_{B+} = \eta_{B-} = \eta_B.\]

The constraints for the fixed analyzers version become:
\begin{eqnarray*}
p^{AA'BB'}_{****} & = & 1, \\
p^{AA'BB'}_{\pm***} = p^{AA'BB'}_{*\pm**} & = & \eta_A, \\
p^{AA'BB'}_{**\pm*} = p^{AA'BB'}_{***\pm} & = & \eta_B, \\
p^{AA'BB'}_{\pm*\pm*} = p^{AA'BB'}_{\pm**\pm} = p^{AA'BB'}_{*\pm\pm*} = p^{AA'BB'}_{*\pm*\pm} & = & \eta_A \eta_B, \\
p^{AA'BB'}_{\pm\pm**} & = & \eta_A^2, \\
p^{AA'BB'}_{**\pm\pm} & = & \eta_B^2, \\
p^{AA'BB'}_{\pm*\pm\pm} = p^{AA'BB'}_{*\pm\pm\pm} & = & \eta_A \eta_B^2, \\
p^{AA'BB'}_{\pm\pm\pm*} = p^{AA'BB'}_{\pm\pm*\pm} & = & \eta_A^2 \eta_B, \\
p^{AA'BB'}_{\pm\pm\pm\pm} & = & \eta_A^2 \eta_B^2.
\end{eqnarray*}

For the removable analyzers version, the constraints become:
\begin{eqnarray*}
p^{AA'A''BB'B''}_{******} & = & 1, \\
p^{AA'A''BB'B''}_{\pm*****} = p^{AA'A''BB'B''}_{*\pm****} = p^{AA'A''BB'B''}_{**+***} & = & \eta_A, \\
p^{AA'A''BB'B''}_{***\pm**} = p^{AA'A''BB'B''}_{****\pm*} = p^{AA'A''BB'B''}_{*****+} & = & \eta_B, \\
p^{AA'A''BB'B''}_{\pm**\pm**} = p^{AA'A''BB'B''}_{\pm***\pm*} = p^{AA'A''BB'B''}_{\pm****+} & = & \eta_A \eta_B, \\
p^{AA'A''BB'B''}_{*\pm*\pm**} = p^{AA'A''BB'B''}_{*\pm**\pm*} = p^{AA'A''BB'B''}_{*\pm***+} & = & \eta_A \eta_B, \\
p^{AA'A''BB'B''}_{**+\pm**} = p^{AA'A''BB'B''}_{**+*\pm*} = p^{AA'A''BB'B''}_{**+**+} & = & \eta_A \eta_B, \\
p^{AA'A''BB'B''}_{\pm\pm****} = p^{AA'A''BB'B''}_{\pm*+***} = p^{AA'A''BB'B''}_{*\pm+***} & = & \eta_A^2, \\
p^{AA'A''BB'B''}_{***\pm\pm*} = p^{AA'A''BB'B''}_{***\pm*+} = p^{AA'A''BB'B''}_{****\pm+} & = & \eta_B^2, \\
p^{AA'A''BB'B''}_{\pm\pm*\pm**} = p^{AA'A''BB'B''}_{\pm\pm**\pm*} = p^{AA'A''BB'B''}_{\pm\pm***+} & = & \eta_A^2 \eta_B, \\
p^{AA'A''BB'B''}_{\pm*+\pm**} = p^{AA'A''BB'B''}_{\pm*+*\pm*} = p^{AA'A''BB'B''}_{\pm*+**+} & = & \eta_A^2 \eta_B, \\
p^{AA'A''BB'B''}_{\pm**\pm\pm*} = p^{AA'A''BB'B''}_{\pm**\pm*+} = p^{AA'A''BB'B''}_{\pm***\pm+} & = & \eta_A \eta_B^2, \\
p^{AA'A''BB'B''}_{*\pm+\pm**} = p^{AA'A''BB'B''}_{*\pm+*\pm*} = p^{AA'A''BB'B''}_{*\pm+**+} & = & \eta_A^2 \eta_B, \\
p^{AA'A''BB'B''}_{*\pm*\pm\pm*} = p^{AA'A''BB'B''}_{*\pm*\pm*+} = p^{AA'A''BB'B''}_{*\pm**\pm+} & = & \eta_A \eta_B^2, \\
p^{AA'A''BB'B''}_{**+\pm\pm*} = p^{AA'A''BB'B''}_{**+\pm*+} = p^{AA'A''BB'B''}_{**+*\pm+} & = & \eta_A \eta_B^2, \\
p^{AA'A''BB'B''}_{\pm\pm+***} & = & \eta_A^3, \\
p^{AA'A''BB'B''}_{***\pm\pm+} & = & \eta_B^3, \\
p^{AA'A''BB'B''}_{\pm\pm+\pm**} = p^{AA'A''BB'B''}_{\pm\pm+*\pm*} = p^{AA'A''BB'B''}_{\pm\pm+**+} & = & \eta_A^3 \eta_B, \\
p^{AA'A''BB'B''}_{\pm\pm*\pm\pm*} = p^{AA'A''BB'B''}_{\pm\pm*\pm*+} = p^{AA'A''BB'B''}_{\pm\pm**\pm+} & = & \eta_A^2 \eta_B^2, \\
p^{AA'A''BB'B''}_{\pm*+\pm\pm*} = p^{AA'A''BB'B''}_{\pm*+\pm*+} = p^{AA'A''BB'B''}_{\pm*+*\pm+} & = & \eta_A^2 \eta_B^2, \\
p^{AA'A''BB'B''}_{\pm**\pm\pm+} = p^{AA'A''BB'B''}_{*\pm*\pm\pm+} = p^{AA'A''BB'B''}_{**+\pm\pm+} & = & \eta_A \eta_B^3, \\
p^{AA'A''BB'B''}_{*\pm+\pm\pm*} = p^{AA'A''BB'B''}_{*\pm+\pm*+} = p^{AA'A''BB'B''}_{*\pm+*\pm+} & = & \eta_A^2 \eta_B^2, \\
p^{AA'A''BB'B''}_{\pm\pm+\pm\pm*} = p^{AA'A''BB'B''}_{\pm\pm+\pm*+} = p^{AA'A''BB'B''}_{\pm\pm+*\pm+} & = & \eta_A^3 \eta_B^2, \\
p^{AA'A''BB'B''}_{\pm\pm*\pm\pm+} = p^{AA'A''BB'B''}_{\pm*+\pm\pm+} = p^{AA'A''BB'B''}_{*\pm+\pm\pm+} & = & \eta_A^2 \eta_B^3, \\
p^{AA'A''BB'B''}_{\pm\pm+\pm\pm+} & = & \eta_A^3 \eta_B^3.
\end{eqnarray*}

From these constraints and the non-negativity of all probabilities, result the following Bell inequalities:
\begin{widetext}
\begin{eqnarray*}
2 \eta_A^2 \eta_B^2 - 4 \eta_A \eta_B \leqslant & S & \leqslant -2 \eta_A^2 \eta_B^2 + 4 \eta_A \eta_B, \\*
-\eta_A^2 \eta_B^2 + \eta_A \eta_B ( \eta_A + \eta_B ) \leqslant & \Delta' & \leqslant 0, \\*
-\eta_A^3 \eta_B^3 + 3 \eta_A^2 \eta_B^2 -3 \eta_A \eta_B \leqslant & \Delta & \leqslant \eta_A^3 \eta_B^3 -
\eta_A^2 \eta_B^2 ( \eta_A + \eta_B ) + \eta_A \eta_B ( \eta_A^2 + \eta_B^2 -2 ( \eta_A + \eta_B ) + 3 ), \\*
& \delta & \leqslant \frac{\eta_A^3 \eta_B^3}{2} - \frac{\eta_A^2 \eta_B^2 ( \eta_A + \eta_B + 3 )}{4} +
\frac{\eta_A \eta_B ( \eta_A^2 + \eta_B^2 -2 ( \eta_A + \eta_B ) + 6 )}{4}.
\end{eqnarray*}

Normalized Bell-quantities, for asymmetric EPRB experiments, are defined as:
\[S_{\text{N}} := \frac{S}{\eta_A \eta_B}, \qquad \Delta_{\text{N}} := \frac{\Delta}{\eta_A \eta_B}, \qquad \delta_{\text{N}} := \frac{\delta}{\eta_A \eta_B},\]
and, consequently, Bell inequalities for the normalized quantities are:
\begin{eqnarray*}
2 \eta_A \eta_B - 4 \leqslant & S_{\text{N}} & \leqslant -2 \eta_A \eta_B + 4, \\*
-\eta_A^2 \eta_B^2 + 3 \eta_A \eta_B -3 \leqslant & \Delta_{\text{N}} & \leqslant \eta_A^2 \eta_B^2 - \eta_A \eta_B ( \eta_A + \eta_B ) +
\eta_A^2 + \eta_B^2 -2 ( \eta_A + \eta_B ) + 3, \\*
& \delta_{\text{N}} & \leqslant \frac{\eta_A^2 \eta_B^2}{2} - \frac{\eta_A \eta_B ( \eta_A + \eta_B + 3 )}{4} +
\frac{\eta_A^2 + \eta_B^2 -2 ( \eta_A + \eta_B ) + 6}{4}.
\end{eqnarray*}
\end{widetext}

\begin{figure}[ht]
\includegraphics[bb = 60 395 540 800, scale = 0.32, clip]{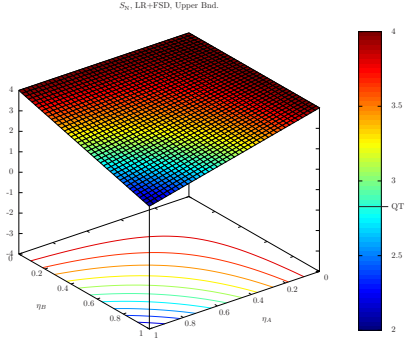}
\caption{Upper bound on Bell-quantity $S_{N}$.\label{SNUB}}
\end{figure}

\begin{figure}[ht]
\includegraphics[bb = 60 395 540 800, scale = 0.32, clip]{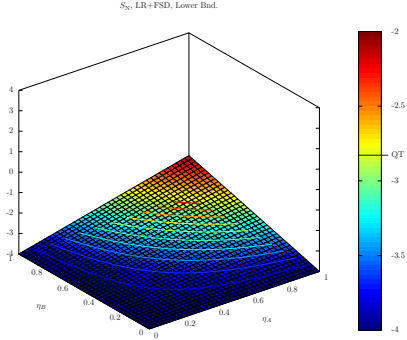}
\caption{Lower bound on Bell-quantity $S_{N}$.\label{SNLB}}
\end{figure}

\begin{figure}[ht]
\includegraphics[bb = 60 395 540 810, scale = 0.32, clip]{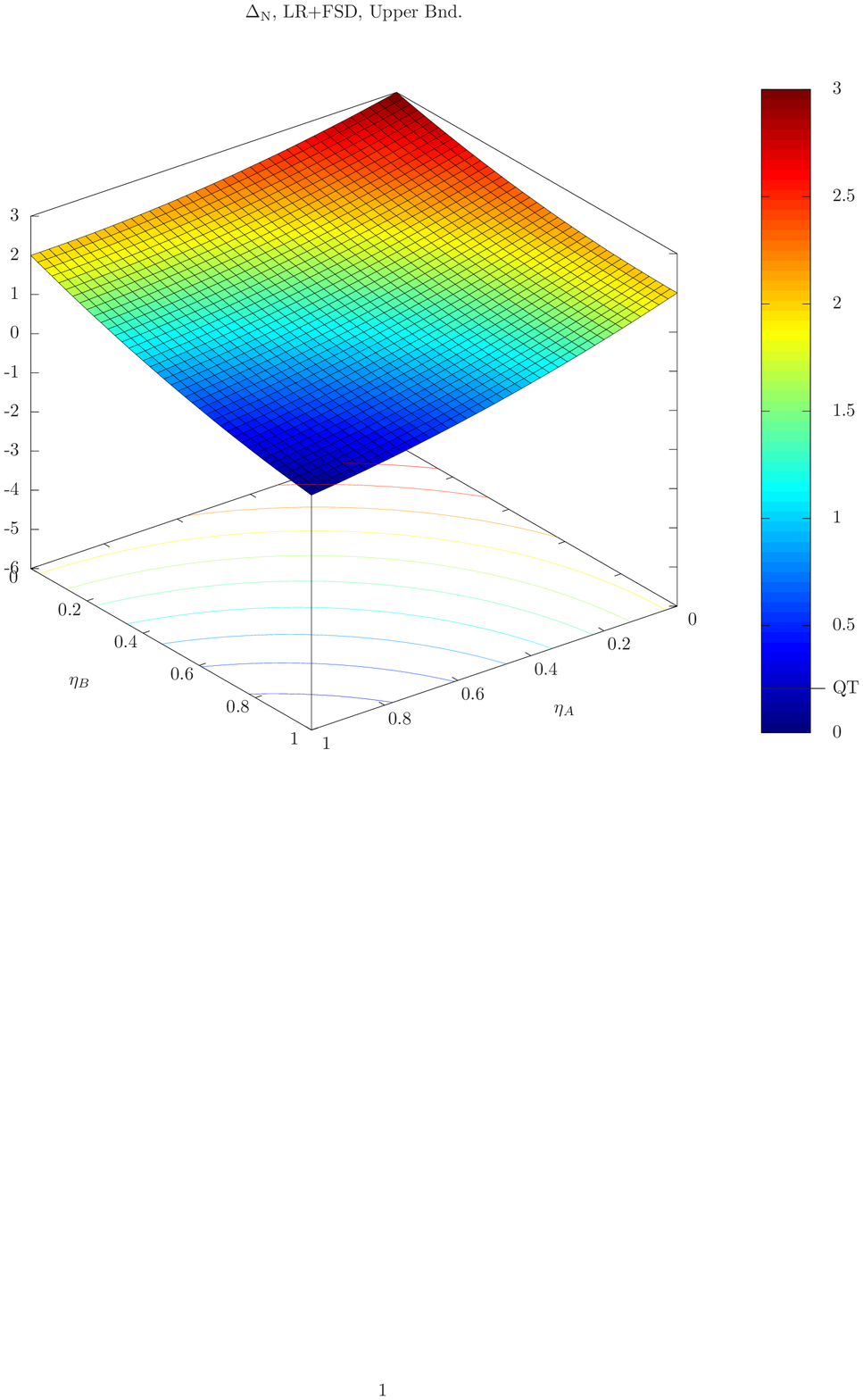}
\caption{Upper bound on Bell-quantity $\Delta_{N}$.\label{DeltaNUB}}
\end{figure}

\begin{figure}[ht]
\includegraphics[bb = 60 395 540 810, scale = 0.32, clip]{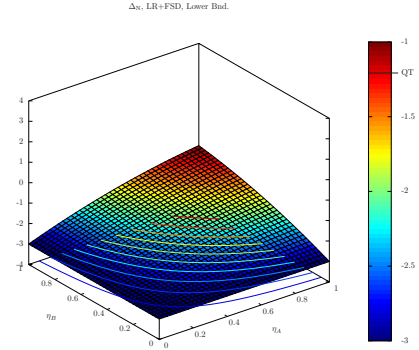}
\caption{Lower bound on Bell-quantity $\Delta_{N}$.\label{DeltaNLB}}
\end{figure}

\begin{figure}[ht]
\includegraphics[bb = 60 395 540 810, scale = 0.32, clip]{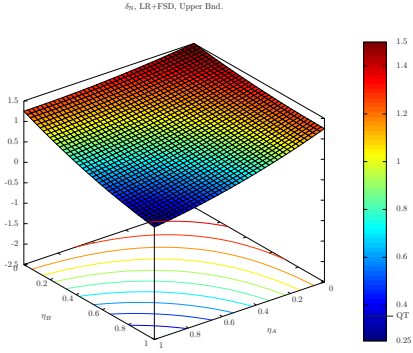}
\caption{Upper bound on Bell-quantity $\delta_{N}$.\label{deltaNUB}}
\end{figure}

The above upper and lower bounds are depicted graphically, for each normalized Bell-quantity, in Figures~\ref{SNUB}~to~\ref{deltaNUB}.

The respective values predicted by quantum theory, (\ref{QTFSB5})-(\ref{QTFSB7}), are marked in the Figures' legends by the QT lines.
\clearpage
Once again, the predictions of quantum theory can be seen to be compatible with the bounds imposed by local-realism and fair-sampling detection as long as $\eta_A$ and $\eta_B$ both lie below variable critical values, a tradeoff existing between detection efficiency on one arm and critical detection efficiency on the other.

For the normalized Bell-quantity $S_{\text{N}}$, compatibility between quantum theory and the conjunction of local-realism and fair-sampling detection exists as long as:
\[\eta_A \eta_B \leqslant 2 - \sqrt{2} \approx 0.5858.\]

Hence, if detection efficiency in one arm increases above $\sqrt{2 - \sqrt{2}} \approx 0.7654$, critical efficiency on the other decreases below this value in inverse proportion.
If detection efficiency became ideal in one arm, critical efficiency on the other could be as low as $0.5858$.

For Bell tests based on the normalized Bell-quantity $S_{\text{N}}$, the criteria for their classification into the three categories mentioned in Section~\ref{Bell theorems} can now be quantified:
\begin{enumerate}
\item
Those in which detection efficiency is lower than $0.7654$ on both arms fall into the category of having inefficient detection on both arms and were discussed in Section~\ref{Inefficient detection on both arms};
\item
Those which have detection efficiency higher than $0.7654$ on one arm but lower than this value on the other are designated as asymmetric tests and will be discussed in the remainder of this Section;
\item
Those which have detection efficiency higher than $0.7654$ on both arms are said to have nearly ideal detection on both arms and will be discussed in Section~\ref{Measurement crosstalk}.
\end{enumerate}

Asymmetric Bell tests \cite{MMBM04,VWSRVSKW06,RHHKVWW08} have involved a trapped ion, on the arm having high detection efficiency, and a photon, on the inefficient one.

Because in realized Bell tests of this type, detection efficiency on the photon arm has been lower than the critical threshold applicable if detection had been ideal on the ion arm, these experiments are inconclusive regarding discrimination between quantum theory and the conjunction of local-realism and fair-sampling detection.
\section{Measurement crosstalk\label{Measurement crosstalk}}
The last decade has seen the development of two-level quantum systems that can be measured and detected with near certainty, known as qubits.

If the EPRB experimental protocol were performed on two such systems, a nearly ideal EPRB experiment would result.

However, these systems have involved ion traps \cite{RKMSIMW01} or superconducting devices under cryogenic conditions \cite{AWBHLNOSWWCM09} which have prevented their physical separation after their initial state has been prepared.

Performing Bell tests using these systems thus closes the problem of detection efficiency but opens a new one.

Because the two systems are placed in close proximity to enable the preparation of their initial state, they remain so during the measurement stage and the possibility exists of the measurement process performed on one qubit to influence the state, and hence the measurement outcome, of the other.

This problem is known as measurement crosstalk \cite{MSSCCOOPM05} and has been measured in several experiments \cite{MSSCCOOPM05,AWBHLNOSWWCM09,BAHLNOSWWSCM10,ACSASLPSTWS10}.

Measurement crosstalk can be modeled by allowing the joint probabilities in Table~\ref{JProbTables} to deviate, by at most the probability of crosstalk, $p^{\text{C}}$, from the respective probabilities that result from Table~\ref{LRJProbTable}.

This means relaxing equalities (\ref{T3toT2}) into inequalities:
\begin{equation}
\forall_{i,j \in \{+,-\}} \colon
\left\{
\renewcommand*{\arraystretch}{1.5}
\begin{array}{ccl}
|p^{AB}_{ij} - p^{AA'BB'}_{i\pm j\pm}| & \leqslant & p^{\text{C}} \\
|p^{AB'}_{ij} - p^{AA'BB'}_{i\pm\pm j}| & \leqslant & p^{\text{C}} \\
|p^{A'B}_{ij} - p^{AA'BB'}_{\pm i j\pm}| & \leqslant & p^{\text{C}} \\
|p^{A'B'}_{ij} - p^{AA'BB'}_{\pm\ i\pm j}| & \leqslant & p^{\text{C}}
\end{array}
\right.. \label{Crosstalk}
\end{equation}

These inequalities imply the following constraints on the marginal probabilities in Table~\ref{JProbTables}:
\begin{eqnarray*}
|p^{AB}_{+\pm} - p^{AB'}_{+\pm}| \leqslant 4 p^{\text{C}}, & \qquad & |p^{AB}_{-\pm} - p^{AB'}_{-\pm}| \leqslant 4 p^{\text{C}}, \\
|p^{A'B}_{+\pm} - p^{A'B'}_{+\pm}| \leqslant 4 p^{\text{C}}, & \qquad & |p^{A'B}_{-\pm} - p^{A'B'}_{-\pm}| \leqslant 4 p^{\text{C}}, \\
|p^{AB}_{\pm+} - p^{A'B}_{\pm+}| \leqslant 4 p^{\text{C}}, & \qquad & |p^{A'B}_{\pm-} - p^{A'B}_{\pm-}| \leqslant 4 p^{\text{C}}, \\
|p^{AB'}_{\pm+} - p^{A'B'}_{\pm+}| \leqslant 4 p^{\text{C}}, & \qquad & |p^{AB'}_{\pm-} - p^{A'B'}_{\pm-}| \leqslant 4 p^{\text{C}}.
\end{eqnarray*}
from which follows that absence of crosstalk necessarily implies apparent locality.

Conversely, lack of apparent locality places a lower bound on the probability of crosstalk.
Should $\Delta p$ be the largest of the above absolute values of differences in marginal probabilities:
\[p^{\text{C}} \geqslant \frac{\Delta p}{4}.\]

From inequalities (\ref{Crosstalk}), together with non-negativity for all probabilities and totals of 1 for all sub-tables in Table~\ref{JProbTables} as well as for Table~\ref{LRJProbTable}, result the following bounds for Bell-quantity $S$, applicable, under crosstalk, to the conjunction of local-realism and ideal detection on both arms:
\[|S| \leqslant \left\{
\renewcommand*{\arraystretch}{1.5}
\begin{array}{lr}
2 + 16\,p^{\text{C}} & \colon 0 \leqslant p^{\text{C}} \leqslant \frac{1}{8} \\
4 & \colon \frac{1}{8} \leqslant p^{\text{C}} \leqslant 1
\end{array}
\right..\]

These bounds remain unchanged if the requirement of apparent locality is added, through inclusion also of equalities (\ref{ApLoc1})-(\ref{ApLoc4}) in the constraints of the parametric linear programming problem.
Just as for locality, apparent locality is a necessary but not sufficient condition for absence of crosstalk.

Measurement crosstalk can thus be seen to require a correction to the ideal Bell inequality (\ref{LRB1}) larger than previously estimated \cite{KK08}.

Quantum theory's predictions for EPRB experiments with ideal detection, (\ref{QTB1}), can be seen compatible with the above bounds for $p^{\text{C}} \geqslant p^{\text{C}}_{\text{c}} = \frac{\sqrt{2}-1}{8} \approx 0.0518$.

To our knowledge, only one Bell test involving two qubits has, to this date, achieved a probability of crosstalk lower than this critical threshold: \cite{AWBHLNOSWWCM09}.
Its results are recalled in Table~\ref{CrosstalkTable}.

\begin{table}[ht]
\caption{Comparison between the results of \cite{AWBHLNOSWWCM09} and local-realism.\label{CrosstalkTable}}
\begin{ruledtabular}
\renewcommand*{\arraystretch}{1.5}
\begin{tabular}{cc}
$p^{\text{C}}_{A \to B}$ & $0.31\%$ \\
$p^{\text{C}}_{B \to A}$ & $0.59\%$ \\
$\bar{p}^{\text{C}}$ & $0.45\% \pm 0.14\%$ \\
\hline
$\hat{S}^{\text{UB}}$ & $2.0720 \pm 0.0224$ \\
$S^{\text{Exper}}$ & $2.0732 \pm 0.0003$ \\
\hline
H. t. $S^{\text{Exper}} = \hat{S}^{\text{UB}}$ & $z = 0.0536$, $\alpha = 0.9573$
\end{tabular}
\end{ruledtabular}
\end{table}

In this test, the largest absolute value of the differences in marginal probabilities was $\Delta p = 0.88\%$ \cite{AWBHLNOSWWCM09} which places a lower bound of $0.22\%$ on the probability of crosstalk, compatible with the measured values.

The measured value of the Bell-quantity, $S^{\text{Exper}}$, can be seen to be not only compatible but actually consistent with the upper bound predicted by local-realism, $\hat{S}^{\text{UB}}$, for the observed probability of crosstalk, $\bar{p}^{\text{C}}$.

From a local-realistic point of view, this agreement is understandable: Since an optimization search was performed on all relevant parameters of this experiment to maximize the measured value of $S$ \cite{AWBHLNOSWWCM09}, maximum use of available crosstalk was achieved.

Bell-like quantities have also been measured in experiments which involve a single quantum system instead of two quantum systems \cite{HLBBR04,BKSSCRH09,SKZBH10,SKDLFLBBRH10}.

In these experiments, two different properties are measured on each prepared quantum system and each measurement is carried out with two different parameter settings.

Since these measurements are performed sequentially in time, the first measurement process may alter the state of the property measured on the second measurement and, consequently, the outcome of the second measurement may be influenced by the choice of parameter adopted for the first.

This is again a situation of measurement crosstalk. In these experiments, even if apparent locality was observed in the measured marginal probabilities, this would not, as shown above, be sufficient to exclude the presence of measurement crosstalk.

Such experiments so fundamentally depart from the EPRB experiment design that they do not measure Bell-quantities, nor constitute tests of local-realism, nor of hypotheses it implies, namely, non-contextuality.
\section{Conclusion\label{Conclusion}}
In this article, realized Bell tests were classified into three categories.

The first, experiments having inefficient detection on both arms, includes, among others, all purely optical Bell tests.
The results from experiments in this category were shown to be compatible with the conjunction of local-realism and fair-sampling detection.

The second, asymmetric Bell tests in which detection is nearly ideal on one arm but inefficient on the other, include all experiments involving an atomic qubit and a photon and their results were also shown to be compatible with the conjunction of local-realism and fair-sampling detection.

Finally, of all Bell tests involving two qubits, which provide nearly ideal detection on both arms but allow measurement crosstalk between them, only one has achieved a sufficiently small probability of crosstalk to be able to discriminate between quantum theory and local-realism and its results were shown to be not only compatible but, actually, in agreement with local-realism.

All published evidence from experimental Bell tests has thus, been shown jointly compatible with local-realism and fair-sampling detection.

More than 75 years since Einstein expressed his belief that a local-realistic theory of quantum phenomena should be possible \cite{EPR35}, and after nearly 40 years in which countless experiments, specifically designed to discriminate between quantum theory and local-realism, have been performed, no evidence still exists that local-realism does not apply to quantum phenomena.
\end{document}